\documentclass{modified}
\usepackage{tikz}
\usetikzlibrary{decorations.pathmorphing}
\usepackage{bm}
\usepackage{bbm}

\newcommand{\jac}{\mathcal{J}}
\newcommand{\thet}{(k(a-(c-U^{(0)}(b))t)}

\newcommand{\eps}{\epsilon}
\newcommand{\defeq}{\overset{\text{def}}{=}}

\newcommand{\thetzero}{(k(a-(c^{(0)}-U^{(0)}(b))t))}

\newcommand{\ssup}[1]{^{(#1)}}

\lefttitle{L. R. Seitz, M. A. Freilich, and N. Pizzo}

\title{The role of Lagrangian drift in the generation of surface waves by wind}

\author{L. R. Seitz \aff{1}, Mara A. Freilich \aff{1,}\aff{2}, and Nick Pizzo\aff{3}}

\affiliation{\aff{1} Division of Applied Mathematics, Brown University, Providence, RI, USA 
\aff{2} Department of Earth, Environmental, and Planetary Science, Brown University, Providence, RI, USA \aff{3} Graduate School of Oceanography, University of Rhode Island, Narragansett, RI, 02882, USA}

\corresau{L.R. Seitz, \texttt{lrseitz@brown.edu}}

\begin{document}
\maketitle

\begin{abstract}
\textbf{Abstract.} A nonlinear stability analysis entirely in the Lagrangian frame is conducted, revealing the fundamental role of the wave-induced mean flow in modifying further wave growth and providing new insight into the classic problem of wave generation by wind. The prevailing theory, a critical-layer resonance mechanism proposed by Miles (1957), has seen numerous refinements; yet, the role of Lagrangian drift -- the velocity a fluid parcel actually experiences -- in wave growth was not understood. Our analysis first recovers the classic Miles growth rate from linear theory before extending it to third order in the wave slope to derive a modified growth rate. The leading-order wave-induced mean flow alters the higher-order instability, manifesting as a suppression of growth with increasing wave steepness for the realistic wind profiles considered. This result is qualitatively consistent with observations. An integral momentum budget suggests that the wave-induced current alters the coupling between the total phase speed and the \emph{total} Lagrangian mean flow at the critical level (as defined in the linear theory), thereby reducing the efficiency of momentum transfer. Notably, this Lagrangian drift is precisely what Doppler-shift based remote sensing of upper ocean currents measure, providing a direct observational pathway to account for this wave-induced feedback in studies of air-sea coupling. More broadly, this approach provides a new methodology for analyzing shear instabilities in general and a direct path towards refining wind-stress parameterizations.
\end{abstract}
\section{Introduction}\label{sec:introduction}
The generation of ocean waves by wind remains a foundational problem in fluid dynamics, though it has evolved considerably since Miles' \citeyearpar{miles1957} pioneering work establishing that instability due to wind shear constitutes a fundamental physical mechanism for wave growth. Subsequent efforts have developed in complementary yet divergent directions: enhancing the understanding of air-sea momentum transfer by detailing the influence of specific factors — such as shear flow in the water \citep{valenzuela1976growth, young2014}, viscous stresses \citep{benjamin1959shearing, miles1959generation}, and airflow dynamics \citep{al1984turbulent, Riley1982, Grare2013, buckley2020surface, cao2021numerical} — beyond Miles' \citeyearpar{miles1957} original framework, while simultaneously pursuing simplified representations of wind stress to incorporate into ocean models \citep{Plant1982-ty, weber2003wave, Drennan2003-ij}. Yet, this body of work has struggled to resolve key discrepancies found in experimental data \citep{sullivan-mcwilliams2010}. 
\par It is well-established that waves influence momentum transfer into the upper ocean, through both wave-induced Reynolds stresses \citep{van1992analytic, miles1993surface} and the leading-order wave-induced current \citep{young2014}. Beyond these interfacial effects, the interaction between surface waves and shear flow can generate Langmuir circulation, which drives an efficient downward redistribution of horizontal momentum and tracers from the surface \citep{thorpe2004, belcher2012, hamlington2014, li2017}. The full role of the wave-induced current in the instability mechanism itself, however, has remained poorly understood. This theoretical gap may help explain phenomena such as the systematic dependence of wave growth rates on mean wave steepness found in laboratory experiments and indicated by observations \citep{peirson-garcia2008}. These considerations lead to the central question of the present work: does the wave-induced mean current play a role in the resonance interaction underlying the physical mechanism behind wave growth? Answering this question also holds the potential to provide a new perspective where the influence of other factors, \emph{e.g.} surface roughness or induced Reynolds stresses, on wave growth is understood through their intrinsic connection to the wave-induced mean flow.
\par In the classic theory, Miles established that the growth of surface gravity waves due to wind can be predicted based on a resonance interaction between the intrinsic frequency $c$ of an initial sinusoidal disturbance on the water surface and the leading-order Eulerian mean velocity $U(z)$ at a critical level $z_c$ in the air. Yet, the wavy disturbance creates a distinction between Eulerian and Lagrangian descriptions of the mean flow, perturbing the mean motion of parcels in the bulk of both the air and the water. This difference between the Lagrangian-averaged velocity and the Eulerian-averaged velocity is the Stokes drift \citep{andrews1978exact}, $\overline{\bm{u}}_L - \overline{\bm{u}}_E=\bm{u}_S.$ In Miles \citeyearpar{miles1957}, $U(z)=\overline{\bm{u}}_E$ at leading order. The question of the role of the wave-induced current in the instability can thus reduce to understanding the role of the full \emph{Lagrangian mean}. This is further motivated by recent findings showing that (i) remote sensing techniques based on observing current-induced shifts in the wave dispersion measure the Lagrangian, rather than Eulerian, mean current \citep{pizzo2023}; (ii) Stokes drift likely contributes significantly to upper ocean shear \citep{lenain2023airborne}; and (iii) including a Stokes drift term when modeling the time evolution of nonlinear surface waves avoids significant errors in the wave celerity \citep{guerin2019}. 
\par Recent advances in Lagrangian theory highlight a promising approach for addressing these questions. These advances include a rigorous asymptotic expansion for the Lagrangian description of steady surface gravity waves \citep{clamond2007lagrangian}, the development of a second-order Lagrangian description of surface gravity wave interactions that captures nonlinear phenomena absent in Eulerian descriptions of the same order \citep{nouguier2015}, clarification of how Lagrangian drift influences the geometry, phase speed, and stability properties of surface waves \citep{pizzo2023}, and the demonstration that Lagrangian drift may be identified with mean Lagrangian momentum density and is determined by vorticity \citep{Blaser2024}. Despite these advances, the role of Lagrangian drift in the instability is \emph{a priori} unclear. Although the Rayleigh equation has been formulated in the Lagrangian frame \citep{bennett2006}, we present the first complete analysis of any hydrodynamic shear instability in the Lagrangian frame. This methodology also addresses the absence of formal Lagrangian analysis in the physical understanding of the instability evolution, which is essentially a Lagrangian parcel argument \citep{Lighthill1962}.
\par While the Eulerian and Lagrangian reference frames are theoretically equivalent, the Lagrangian framework offers distinct advantages for analyzing wave-induced currents. Prior studies used Lagrangian equations of motion to formulate evolution equations for wave-induced mass transport in the ocean, leading to alternative models for wave growth, albeit in a context different from the Miles mechanism \citep{weber1983, Jenkins1986-ov, Weber1993-fy}. A Lagrangian approach thus holds the potential to resolve both the theoretical question of how Lagrangian drift influences wave generation and the practical question of which reference frame is best for wind stress estimation. Moreover, this perspective can reveal physical interpretations that are hidden in an Eulerian view.
\par In $\S$2, by asymptotically expanding Lagrangian trajectories in terms of permanent progressive monochromatic waves, we demonstrate that it is possible to conduct the stability analysis as in \cite{miles1957} and \cite{young2014} in the Lagrangian frame. While our analysis recovers the classic growth rate (of the linear theory), additional insights about its spatial dependence and the shape of the critical layer emerge due to the Lagrangian perspective. In $\S$3, we extend the analysis through third order in the wave slope to derive a modified instability growth rate, showing how it is altered by the (leading-order) wave-induced mean flow. In addition, we derive the formula for the curvature of the free surface in Lagrangian coordinates and the corresponding higher-order correction to the growth rate due to surface tension, extending the classical theory to account for nonlinear wave-mean flow interactions in the presence of capillarity. Lastly, in $\S$4, we discuss the physical insights arising from our Lagrangian analysis. This includes an integral momentum budget, which reveals that the interaction between the \emph{total} phase speed $c$ (summing contributions at all orders of the wave slope) and the \emph{total} Lagrangian mean flow $U$ plays a critical role in the net momentum input from the wind. This confirms that the instability is dependent on the complete, wave-altered velocity field that fluid parcels experience, consistent with qualitative arguments from prior work. Finally, we discuss the broader applicability of this Lagrangian framework to other shear instabilities, and the consequences for estimating air-sea momentum transfer.
\section{Formulation of the linear stability problem in Lagrangian coordinates}\label{sec:2}
\begin{figure}[h]
\centering
\includegraphics[width=0.75\textwidth]{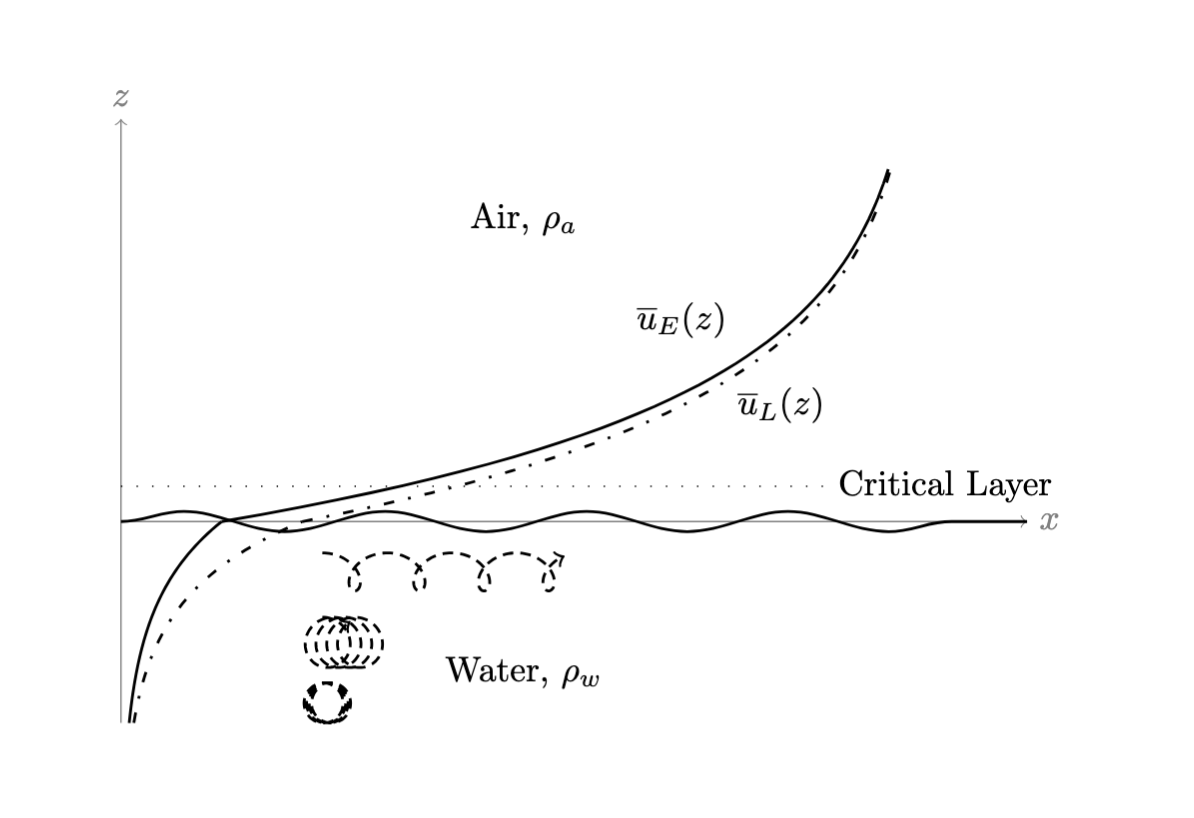}
\caption{The setup of the Miles instability in Lagrangian coordinates, adapted from \cite{young2014}. The Lagrangian mean  velocity profile, $\overline{u}_L(z)$ (identified as $U$ in \eqref{def:monochromatic}), is shown alongside the Eulerian mean, $\overline{u}_E(z)$. While the two profiles are identical at leading order, they differ at higher order due to the surface wave; note that $\overline{u}_L$ is more naturally a function of the particle label $b$ but is shown in terms of $z$ for comparison. Example trajectories (dashed), with the same initial $x$-coordinate, are in the water. As the Lagrangian mean velocity approaches zero, the orbits of particles become closed circles. } 
    \label{fig:1}
\end{figure}
\subsection{The Lagrangian governing equations}
\par In the Lagrangian frame, fluid particles (or parcels) are tracked using fixed labels $(a,b)$ and time $t$ as independent variables. Time derivatives are taken with respect to fixed particle labels $(a,b)$. Trajectories are then described by $\{(x(a,b,t), z(a,b,t))\}_{t \in \mathbb{R}^+}$. The mapping between label space $(a,b)$ and physical space $(x,z)$ is invertible, as no two particles can occupy the same physical location simultaneously. This implies a nonzero Jacobian:
\begin{equation}\label{def:jacobian}
    \mathcal{J} \defeq \frac{\partial(x,z)}{\partial(a,b)}= x_az_b-x_bz_a\neq 0,
\end{equation}
where $x_a$ denotes the partial derivative of the $x$-coordinate with respect to particle label $a$, and similarly for the other derivatives. While this mapping is generally time-dependent, our choice of labels based on particle locations at time $t=0$ ensures that for an incompressible fluid, the Jacobian is constant along a trajectory, \emph{i.e.}, $\dot{\mathcal{J}}=0$. Beyond these constraints, there remains considerable flexibility in particle label assignment.
\par We choose particle labels such that the free surface corresponds to $b=0$. We assign labels based on particle positions at time $t=0$, with label $a$ following the $x$-axis. Our two-fluid system consists of water ($b<0$) and air $(b>0)$. This Lagrangian formulation offers a significant geometric advantage: the time-varying free surface can be described as a simple, fixed plane in label space rather than an additional function $z=\eta(x)$. The conservation of particle labels implies that particles at the surface remain at the surface for all time. This approach directly circumvents the well-known difficulties of defining physically meaningful averages in the Eulerian frame near an undulating interface, a challenge that has motivated the development of analogous coordinate transformations in experimental and numerical work (\emph{e.g.}, \citealt{buckley2016};  \citealt{sullivan2014}). Furthermore, this formulation has a significant analytical advantage in that Lagrangian approximations are generally more accurate and converge faster than Eulerian ones of the same order. For moderately steep waves, for instance, an $n$th order Lagrangian approximation of the surface can match the accuracy of an Eulerian approximation of order $n+2$ \citep{clamond2007lagrangian}. 
\par The Euler equations in Lagrangian coordinates are given by \citep[Art.15]{lamb1924hydrodynamics}
\begin{subequations}
    \begin{align}
        \mathcal{J}\ddot{x} + \frac{1}{\rho}(p_az_b - p_bz_a)&=0 \label{eq:euler-in-lagrangian-x}\\
        \mathcal{J}\ddot{z} + \frac{1}{\rho}(p_bx_a-p_ax_b + \mathcal{J} g)&=0 \label{eq:euler-in-lagrangian-z}
    \end{align}
\end{subequations}
where $p$ is pressure, $\rho$ is density, and $g$ is the gravitational constant. Since particle density remains constant along trajectories, we take $\rho=\rho_0$ as constant. As in the classical stability analysis \citep{miles1957}, viscosity effects are ignored. Lastly, the vorticity is defined as
\begin{equation}\label{def:vorticity}
    \Gamma \defeq v_x-u_y = \frac{1}{\mathcal{J}}(\dot{x}_a x_b - \dot{x}_bx_a + \dot{z}_az_b - \dot{z}_bz_a).
\end{equation}
Vorticity is always conserved, so that $\dot{\Gamma}=0$. For irrotational flows, $\Gamma=0$.
\subsection{Asymptotic expansions}
\par We consider permanent, progressive, monochromatic, spatially-periodic, two-dimensional waves progressing with velocity $c$ \citep{clamond2007lagrangian, pizzo2023}. The coordinates $(x,z)$ are expanded as 
\begin{equation}\label{def:monochromatic}
    x= a +U(b)t + \sum_{n=1}^\infty x_n(b)\sin (\theta_n), \quad
    z = b +z_0(b) +  \sum_{n=1}^\infty z_n(b)\cos(\theta_n), 
\end{equation}
where $\theta_n\defeq nk(a-(c-U(b))t)$, $k$ is the wavenumber and $U$ is the Lagrangian mean flow (and may be identified with $\overline{u}_L$ evaluated in Lagrangian coordinates). The mean surface level is $z_0(0)\defeq \langle z\rangle|_{b=0}$ where $\langle \cdot \rangle=\frac{k}{2\pi}\int_0^{2\pi/k} \cdot \;\mathrm{d}a$ denotes phase averaging. Pressure is expanded as
\begin{equation}\label{eq:pressure expand}
    p = p_0 - \rho_0gb + \sum_{n=1}^\infty p^{(n)},
\end{equation}
where $p_0$ denotes the constant background pressure, so that $p_0=p_{air}$ in the air and $p_0=p_w$ in the water.  
As in \cite{pizzo2023}, the pressure can be solved as 
\begin{equation}\label{eq:pressure}
    p=\rho_0\left(-gz-\frac{(\dot{x}-c)^2 + \dot{z}^2}{2} + \int_{-\infty}^b (c-U)\Gamma \mathcal{J} \,\text{d}\beta + f(a,t)\right),
\end{equation}
where $f(a,t)$ is a constant of integration. 
\par Imposing ordering in \eqref{def:monochromatic}, through third order in the wave slope $\epsilon$, 
\begin{subequations}
\begin{align}
    x&= a +U\ssup{0}(b)t + \eps x_1(c,k;b)\sin (\theta_1)+ \eps^2 U\ssup{2}(c,k;b)t+\eps^2 x_2(c,k;b) \sin (\theta_2)   \nonumber \\
    &\quad\quad+ \eps^3 U\ssup{3}(c,k;b)t + \eps^3 x_3(b)\sin(\theta_3) + O(\eps^4) \label{def:monochromatic-asymp-x}\\ 
    z &= b  +  \eps z_1(c,k;b) \cos(\theta_1) + \eps^2z_0\ssup{2} + \eps^2z_2(b)\cos(\theta_2)+ \eps^3z_0\ssup{3} \nonumber\\
    &\quad\quad+ \eps^3 z_3(b)\cos(\theta_3) +O(\eps^4)\label{def:monochromatic-asymp-z} \\ 
    p &= p_0 - \rho_0gb + \eps p^{(1)} + \eps^2 p^{(2)} + \eps^3 p\ssup{3}  + O(\eps^4)\label{def:monochromatic-asymp-p}
\end{align}
\end{subequations}
where the total Lagrangian mean flow has been written as $U(b) = \sum_{n=0}^\infty \eps^n U\ssup{n}(b)$, so $U\ssup{0}(b)$ denotes the leading-order portion and $U\ssup{2}(b)$ is associated with the Stokes drift. The terms $z_0\ssup{2}$ and $z_0\ssup{3}$ in $z$ are to ensure $\langle zx_a|_{b=0}\rangle =0$ (note that $z_0\ssup{0}, z_0\ssup{1}=0$). Additionally, the phase speed is expanded as $c=c\ssup{0}+\eps c\ssup{1}+\eps^2c\ssup{2}+O(\eps^3)$. The phase speed is also expanded in this manner in \cite{clamond2007lagrangian}, and this expansion makes it possible to find the modified growth rate in $\S$\ref{sec:modified-growth-rate}. 
\subsection{Instability growth rate and critical level in Lagrangian coordinates}
The linear stability analysis, conducted in the Lagrangian frame, recovers the established Miles instability growth rate. This approach, however, offers a more intuitive understanding of the critical level's geometry, showing it is a wavy surface that follows the interface, which is less apparent in a traditional Eulerian analysis. To calculate the instability growth rate, we first formulate the Rayleigh equation in Lagrangian coordinates, which requires a change of variables:
\begin{subequations}
\begin{align}
x\ssup{1}(a,b,t) &= A(a+U\ssup{0}(b)t, b, t), \;\; A(a,b,t) = \xi(b)\sin(k(a-c\ssup{0}t)) \label{eq:change-of-vars-x}, \\
z\ssup{1}(a,b,t) &= B(a+U\ssup{0}(b)t, b, t), \;\; B(a,b,t)= \frac{\varphi(b)}{k(c\ssup{0}-U\ssup{0})}\cos(k(a-c\ssup{0}t)). \label{eq:change-of-vars-z} 
\end{align}
\end{subequations}
\par The Rayleigh equation depends only on $b$ and is derived from the linear approximations of \eqref{eq:euler-in-lagrangian-x}-\eqref{eq:euler-in-lagrangian-z} and mass continuity (see Appendix \ref{appA}) as
\begin{equation}\label{eq:rayleigh-lagrangian}
\varphi_{bb}- \left(k^2 - \frac{\frac{d^2}{db^2}U\ssup{0}}{c\ssup{0}-U\ssup{0}}\right)\varphi = 0.
\end{equation}
\par We will use the Rayleigh equation after deriving a dispersion relation from the linearized dynamic boundary condition. In Lagrangian coordinates, the linearized dynamic boundary condition at the air-sea interface is given by 
\begin{equation}\label{eq:linearized-dynamic-boundary}
    p^{(1)}(a,0^+,t)-p^{(1)}(a,0^-,t)=T z_{aa}^{(1)}(a,0,t),
\end{equation}
where $T$ is the coefficient of surface tension. 
\par Denote $\tilde{a} \defeq a+U_s\ssup{0}t$, where $U_s\ssup{0}\defeq U\ssup{0}(b=0)=U\ssup{0}(b=0^+,0^-)$.  In the new variables \eqref{eq:change-of-vars-z}, using \eqref{eq:pressure}, \eqref{eq:linearized-dynamic-boundary} reduces to
\begin{equation}\label{eq:disp-relation-changed-var-lin-1}
\begin{split}
    p^{(1)}(\tilde{a},0^+,t)&-p^{(1)}(\tilde{a},0^-,t)= -\rho_{air}gB(\tilde{a},0^+,t)+\rho_{w}gB(\tilde{a},0^-,t) \\
    &+ (c\ssup{0}-U\ssup{0}_s)(\rho_{air}B_b(\tilde{a},0^+,t) - \rho_{w}B_b(\tilde{a},0^-,t)).
\end{split}
\end{equation} 
Define the constants 
\begin{equation}\label{def:normalized-parameters}
\eps_\rho \defeq \frac{\rho_{air}}{\rho_{air}+\rho_w}, \quad \gamma\defeq \frac{T}{\rho_{air}+\rho_w}, \quad \text{and} \quad  g'\defeq g\frac{\rho_w-\rho_{air}}{\rho_{air}+\rho_w},
\end{equation}
and the functions 
\begin{subequations}
\begin{align}
\Xi_{air}(c\ssup{0},k) &\defeq -\frac{\varphi_b(c\ssup{0},k;0^+)}{\varphi(c\ssup{0},k;0)}, \quad \Xi_{w}(c\ssup{0},k) \defeq \frac{\varphi_b(c\ssup{0},k;0^-)}{\varphi(c\ssup{0},k;0)}, \text{ and}
\\
S&\defeq (1-\eps_\rho)\frac{d}{db}U\ssup{0}(0^-)-\eps_\rho \frac{d}{db}U\ssup{0}(0^+).
\end{align}
\end{subequations}
After substituting \eqref{eq:linearized-dynamic-boundary} into \eqref{eq:disp-relation-changed-var-lin-1} and some algebra, we obtain the dispersion relation
\begin{equation}\label{eq:disp-relation-changed-var-lin-fin}
(\eps_\rho \Xi_{air} + (1-\eps_\rho)\Xi_w)(c\ssup{0}-U_s\ssup{0})^2+S(c\ssup{0}-U_s\ssup{0})-g'-\gamma k^2 =0.
\end{equation}
As in \cite{young2014}, we rewrite \eqref{eq:disp-relation-changed-var-lin-fin} as 
\begin{equation}\label{eq:disp-relation-lin-abst-1}
    \mathcal{D}_0(c\ssup{0},k) + \eps_\rho\mathcal{D}_1(c\ssup{0},k) = 0
\end{equation}
where 
\begin{equation}\label{eq:disp-relation-lin-abst-defs}
    \begin{split}
        \mathcal{D}_0(c\ssup{0},k) &\defeq \Xi_w(c\ssup{0},k)(c\ssup{0}-U_s\ssup{0})^2 + S_0(c\ssup{0}-U_s\ssup{0})-g-\gamma k^2, \\ 
    \mathcal{D}_1(c\ssup{0},k)&\defeq (\Xi_{air}(c\ssup{0},k)-\Xi_w(c\ssup{0},k))(c\ssup{0}-U_s\ssup{0})^2 + S_1(c\ssup{0}-U_s\ssup{0}) + 2g, \\
        S_0 &\defeq \frac{d}{db}U\ssup{0}(0^-), \; S_1\defeq-\frac{d}{db}U\ssup{0}(0^-)-\frac{d}{db}U\ssup{0}(0^+).
    \end{split}
\end{equation}
The solution to \eqref{eq:disp-relation-lin-abst-1} will be of the form $c\ssup{0}(k,\eps_\rho)$, so we expand 
\begin{equation}\label{eq:c-expansion}
c\ssup{0}(k,\eps_\rho)=c_0\ssup{0}(k) + \eps_\rho c_1\ssup{0}(k) +O(\eps_\rho^2).
\end{equation}
We then obtain the leading-order balance
\begin{equation}\label{eq:leading-order-balance-disp-lin}
\mathcal{D}_0(c_0\ssup{0}(k),k)=0,
\end{equation} which only involves flow in the water, so does not itself affect the Miles instability. The Miles instability results from a critical level in the air, $b_c>0$, such that
$c_0\ssup{0}(k) = U\ssup{0}(b_c).$
At next order in $\eps_\rho$, using a Taylor expansion in terms of $c\ssup{0}$, we can solve
\begin{equation}\label{eq:c1-linear}
c_1\ssup{0} = -\frac{\mathcal{D}_1(c_0\ssup{0},k)}{\partial_{c\ssup{0}}\mathcal{D}_0(c\ssup{0},k)}.
\end{equation}
\par The instability growth rate is that of the unstable solutions to the eigenvalue problem. Notice the Rayleigh equation \eqref{eq:rayleigh-lagrangian} becomes singular when $U\ssup{0}=c\ssup{0}$; to avoid this singularity, we introduce a complex phase speed 
\begin{equation}\label{eq:complex-phase-speed}
c\ssup{0}=c_r\ssup{0}+ic_i\ssup{0},
\end{equation}
where $U\ssup{0}(b_c)=c_r\ssup{0}$ and $c_i\ssup{0} \ll c_r\ssup{0}$. The form of the perturbation in \eqref{eq:change-of-vars-z} grows in time according to $kc_i\ssup{0}$, so that the instability growth rate (in this linear analysis) is $\omega = k \text {Im } c\ssup{0}$. Noting that $c_0\ssup{0}\in \mathbb{R}$, we approximate the growth rate $\omega$ as 
 \begin{equation}\label{def:linear-growth-rate}
     \omega = \eps_\rho k\text{Im } c_1\ssup{0}.
 \end{equation} We then compute that
\begin{equation}\label{eq:imc1-2}
\text{Im } c_1\ssup{0} = - \frac{\text{Im }\mathcal{D}_1}{\text{Re } \partial_{c\ssup{0}}\mathcal{D}_0},
\end{equation}
where (see Appendix \ref{appA})
\begin{equation}\label{eq:growth-rate-linear}
\text{Im } \mathcal{D}_1 = (c_0\ssup{0}-U_s\ssup{0})^2\pi \frac{\frac{d^2}{db^2}U\ssup{0}_c}{\Big|\frac{d}{db}U\ssup{0}_c\Big|}\frac{|\varphi_c|^2}{|\varphi_s|^2}.
\end{equation}
Here, $U_c\ssup{0}\defeq U\ssup{0}(b_c)$, $\varphi_c\defeq \varphi(b_c)$, and $\varphi_s\defeq \varphi(0)$.
\begin{figure}[t]
    \centering
    \includegraphics[width=\linewidth]{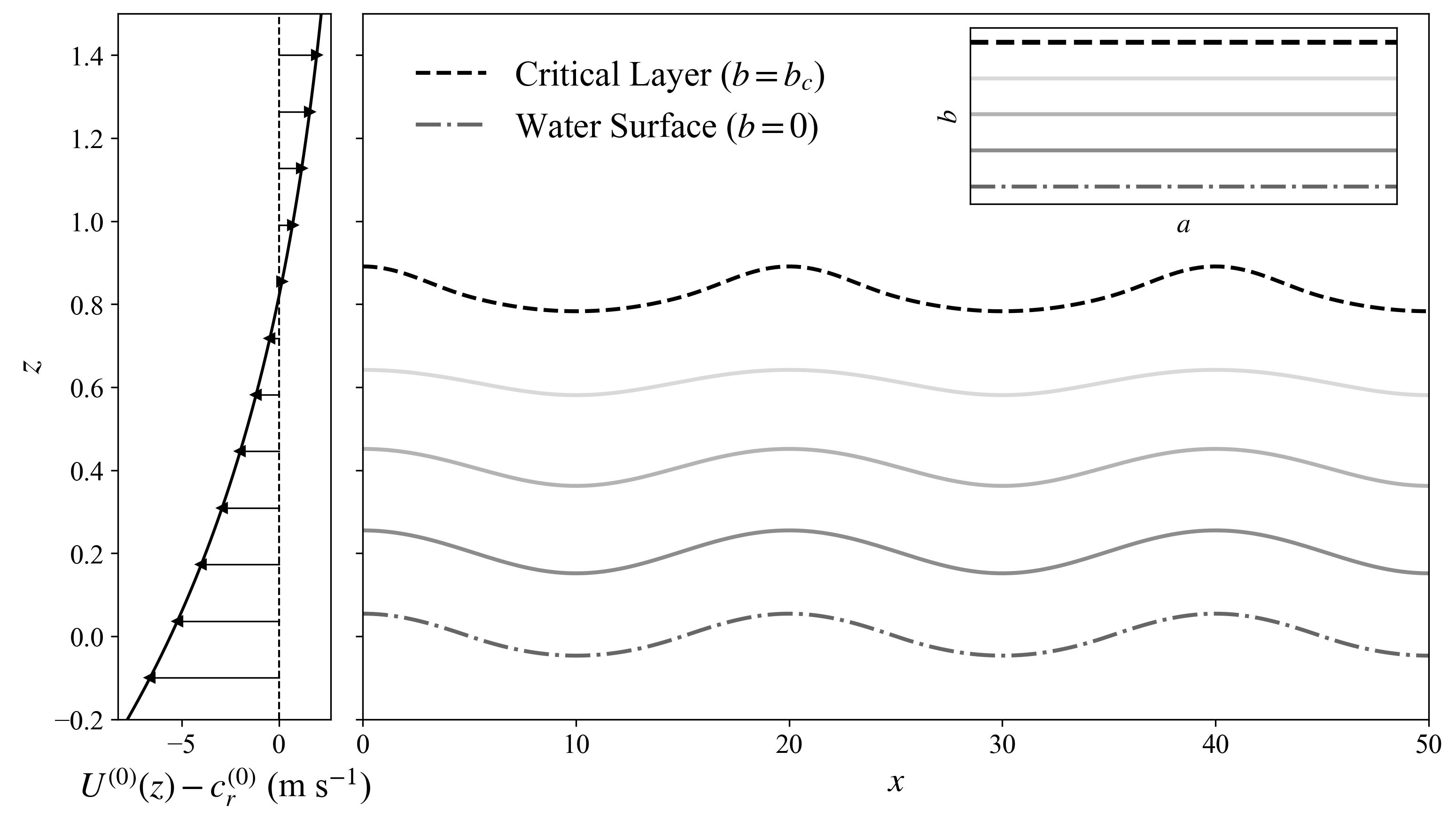}
    \caption{The critical layer defined by  $b=b_c$, the surface defined by $b=0$, and intermediate lines of constant $b$, shown in Eulerian coordinates (the $x-z$ plane) by using the transformations \eqref{eq:inverted-a-linear} and \eqref{eq:inverted-b-linear}. These levels of constant $b$ are also shown in the Lagrangian frame (the $a-b$ plane) in the inset, illustrating the simple representation of the wavy critical level when considering the instability in Lagrangian coordinates. The relative wind is shown on the left. Here, the background wind is the double exponential profile described in $\S$\ref{subsec:double-exp} with $U_s\ssup{0}=0$ (the original case Miles considered, together with $\gamma=0$), $\rho_{air} = 1.25 \ \mathrm{kg \ m^{-3}}$, $\rho_w = 1025 \ \mathrm{kg \ m^{-3}}$, $h_{air} = 1 \ \mathrm{m}$, and $h_w = 0.54 \ \mathrm{cm}$. In this case, $c\ssup{0}_r =(g/k)^{\frac{1}{2}}$, where a wavelength of $k=2\pi / 20$ and a wave slope of $\epsilon=0.1$ were selected (arbitrarily) for visualization. This picture would evolve in time according to \eqref{eq:inverted-b-linear}; here $t=0$ is shown.}
    \label{fig:2}
\end{figure}
\par By inverting $x=x(a,b)$ and $z=z(a,b)$ iteratively at each order of $\eps$, using \eqref{def:monochromatic-asymp-x} and \eqref{def:monochromatic-asymp-z}, we can write $a$ and $b$ in terms of $x$ and $z:$
\begin{subequations}
    \begin{align}
        a(x,z) &= x-U\ssup{0}(z)t-\eps\xi(z)\sin(k(x-c\ssup{0}t)), \label{eq:inverted-a-linear}\\ 
        b(x,z) &= z-\eps\frac{\varphi(z)}{k(c\ssup{0}-U\ssup{0}(z))}\cos(k(x-c\ssup{0}t)). \label{eq:inverted-b-linear}
    \end{align}
\end{subequations}
\par We thus recover the known growth rate for the Miles instability, but in the Lagrangian frame; $U_c\ssup{0}$ and $\varphi_c$ are functions of $b$ rather than $z$, but to leading order $U\ssup{0}(b)=U\ssup{0}(z)$ and $\varphi(b)=\varphi(z)$, per \eqref{eq:inverted-a-linear} and \eqref{eq:inverted-b-linear}. However, the representation of the growth rate \eqref{def:linear-growth-rate} given by \eqref{eq:imc1-2} and \eqref{eq:growth-rate-linear} differs from the Eulerian representation because the arguments of $U\ssup{0}$ and $\varphi$ should be evaluated at the {perturbed} $z$-coordinate described by the right-hand side of \eqref{eq:inverted-b-linear}, rather than simply $z$. The critical level, defined by a constant $b$ (or equivalently, perturbed $z$), is therefore ``wavy," following the shape of the interface. Indeed, \eqref{eq:inverted-b-linear} shows that a line of constant $b$ is not one of constant physical height $z$, but rather a wavy surface oscillating in $x$ (Figure \ref{fig:2}). The growth rate also varies across the interface, since $\varphi_c$ and $U_c\ssup{0}$ in \eqref{eq:growth-rate-linear} occur at a constant level $b_c$, which is {not} a constant height $z_c$ per \eqref{eq:inverted-b-linear}. The waviness of the critical layer, stated by \cite{Lighthill1962} and apparent through the use of orthogonal coordinates in \cite{benjamin1959shearing}, emphasizes that the critical height $z_c$ is better regarded as height from the {actual}, rather than mean, water surface. 
\par This explicit description of the geometry of the critical level is one advantage of the Lagrangian analysis. A more significant outcome of this analysis, however, is the Lagrangian frame's capacity to show how the wave-induced mean flow modifies the instability growth rate, a task that requires extending the stability analysis to higher order. 
\section{Modification to instability growth rate by the wave-induced mean flow}\label{sec:modified-growth-rate}
\subsection{Nonlinear stability analysis}\label{subsec:nonlinear-stab-analysis}
Computing the modification to the growth rate by the wave-induced current requires considering the system through third order. This is similar to other nonlinear stability analyses, \emph{e.g.} \cite{nayfeh1972}, but in the Lagrangian frame, we expand in terms of monochromatic waves \eqref{def:monochromatic} and expand the phase speed and mean velocity (\emph{e.g.}, \citealt{clamond2007lagrangian}) and not the independent variables $a$, $b$, or $t$. In the higher-order analysis, we first consider the special case in which $\gamma=0$ in \eqref{def:normalized-parameters}, as in Miles' original analysis. The full details of the higher-order analysis are provided in Appendix \ref{appB}. While it turns out that $c\ssup{1}=0$ due to the second-order conservation of vorticity, $c\ssup{2}$ is described by the dispersion relation
\begin{equation}\label{eq:c2-disp-relation-no-gamma}
    \rho_{air}(F(0^+)+c\ssup{2}G(0^+))-\rho_w(F(0^-)+c\ssup{2}G(0^-))=0,  
\end{equation}
where 
\begin{subequations}
\begin{align}
    F(b) &\defeq \varphi'(b) \left(\frac{2}{k}\Big(U\ssup{2}(b) + k(c\ssup{0}-U_s\ssup{0})x_2(b)\Big) \right)- 2k(c\ssup{0}-U_s\ssup{0})\varphi_s z_2(b) \nonumber \\
&  + \varphi_s\left(   \frac{1}{k}\left(\frac{d}{db}U\ssup{0}(b)\left( \frac{3U\ssup{2}(b)}{c\ssup{0}-U_s\ssup{0}}+ 2kx_2(b)\right) +  \frac{d}{db}U\ssup{2}(b)\right)\right),  \label{eq:disp-third-order-def-1} \\
G(b) &\defeq -\frac{1}{k(c\ssup{0}-U_s\ssup{0})} \left( 2\varphi'(b)(c\ssup{0}-U_s\ssup{0}) + 3\varphi_s\frac{d}{db}U\ssup{0}(b) \right).
\end{align}
\end{subequations}
\par In contrast to our approach to obtain the linear growth rate, since the dispersion relation \eqref{eq:c2-disp-relation-no-gamma} is only linear in $c\ssup{2}$, we obtain a relatively simple analytical solution without using an asymptotic expansion in $\eps_\rho$:
\begin{equation}\label{eq:c2-sol}
c\ssup{2} = -\frac{\rho_{air}F(0^+)-\rho_wF(0^-)}{\rho_{air}G(0^+)-\rho_wG(0^-)}.
\end{equation}
 Additionally, under the mild assumption that derivatives of $x_n, z_n, n\geq 1$ introduce a constant coefficient of $n$ (as in a Gerstner or Stokes wave), an analytical formula for $x_2$ and $z_2$ in terms of the known lower order quantities $c\ssup{0}$, $U\ssup{0}$, and $z_1$ is possible,
\begin{equation}\label{eq:x2z2intermsofz1} kx_2=z_2' \text{ and } 4z_2'+z_1''z_1-(z_1')^2=0. 
\end{equation}
The leading-order wave-induced mean flow $U\ssup{2}$ can also be expressed in terms of known lower-order quantities (see Appendix \ref{appE}), 
\begin{equation}\label{def:u2}
    U\ssup{2}=\frac{1}{4}(c^{(0)}-U^{(0)})\frac{d^2}{db^2}\left(\frac{\varphi^2}{k^2(c\ssup{0}-U\ssup{0})^2}\right).
\end{equation}
 Thus \eqref{eq:c2-sol} is entirely in terms of known quantities, since we can also use the linear growth rate when considering $c\ssup{0}$, \eqref{def:linear-growth-rate}. However, it may not be computable in general, as $x_2$ may be an indefinite integral that cannot be computed. Nonetheless, \eqref{eq:c2-sol} readily gives an exact solution for the modified growth rate for many relevant background wind profiles $U\ssup{0}$.  Regardless of the ability to compute an exact solution, the dispersion relation clearly indicates that the leading-order growth-rate modification will be directly affected by the wave-induced mean flow, due to the appearance of $U\ssup{2}$ throughout, as well as its shear. 
 \subsection{Growth rates with respect to the double exponential profile}\label{subsec:double-exp}
 \begin{figure}[t]
\begin{center}
\includegraphics[width=0.9\linewidth]{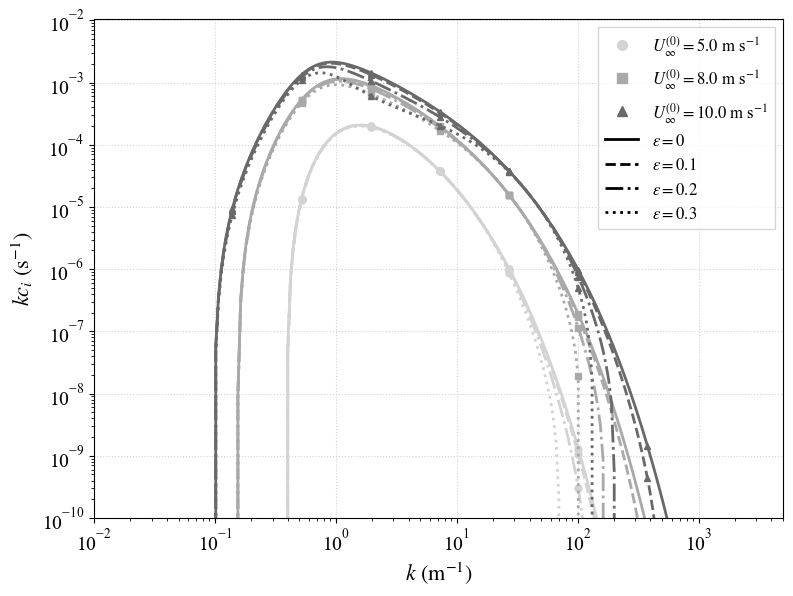}
\end{center}
\caption{The modified growth rate as a function of wavenumber $k$, for four wave slopes $\eps$ and three values of $U_\infty^{(0)}$ (which increases the background shear). The linear growth rate ($\eps=0$, solid) computed in the Lagrangian frame equals the known result computed in the Eulerian frame, while the modified growth rates, dependent on the leading-order wave-induced mean flow ($\eps \neq 0$), show that increased wave slope combined with increased background shear can lead to a significant suppression of the instability at high wavenumbers. For larger values of $U_\infty^{(0)}$, the modification can also significantly reduce the peak growth rate. Growth rates were computed for the double-exponential profile with parameters: $U\ssup{0}_s=0$, $\rho_{air} = 1.25 \ \mathrm{kg \ m^{-3}}$, $\rho_w = 1025 \ \mathrm{kg \ m^{-3}}$, $h_{air} = 1 \ \mathrm{m}$, and $h_w = 0.54 \ \mathrm{cm}$ (as in \cite{young2014}). The values of $\eps$ chosen are based on observational data, \textit{e.g.} those in \cite{peirson-garcia2008}. The linear growth rate shown here is the asymptotic solution, but it exactly aligns with the numerical solution to \eqref{eq:disp-relation-changed-var-lin-fin} (c.f. \cite{young2014}).} 
    \label{fig:3}
\end{figure}
To illustrate the impact of the modified growth rate, we consider an explicit base-state velocity profile 
\begin{equation}\label{eq:double-exponential}
    U\ssup{0}(b) = \begin{cases} U_\infty^{(0)} - (U_\infty^{(0)} - U_s\ssup{0})e^{-b/h_{air}} &\text{ if } b>0, \\ U_s\ssup{0}e^{b/h_w} &\text{ if } b<0.\end{cases}
\end{equation}
This double exponential profile is defined by the scale heights in the air and water ($h_{air}$ and $h_w$) and the velocities at the top of the boundary layer and at the surface ($U_\infty\ssup{0}$ and $U_s\ssup{0}$).
For a given profile, the Rayleigh equation has an exact solution 
\begin{equation}
\label{eq:double-exp-sol}
\varphi(c,k;b) = e^{-k|b|} \frac{F(\alpha_j, \beta_j, 2\kappa_j+1, X_j e^{-|b|/h_j})}{F(\alpha_j, \beta_j, 2\kappa_j+1, X_j)}, 
\end{equation}
where the parameters to the Gaussian hypergeometric function $F(a,b,c,\xi)$ are defined
\begin{equation}
(\alpha_j, \beta_j, \kappa_j, X_j) = \begin{cases}
    \left(\kappa_{air}-\sqrt{1+\kappa_{air}^2}, \kappa_{air}+\sqrt{1+\kappa_{air}^2}, kh_{air}, \frac{U_\infty^{(0)}-U_s\ssup{0}}{U_\infty^{(0)} -c\ssup{0}}\right) & \text{for } j=air, \\[1.2em]
    \left(\kappa_w-\sqrt{1+\kappa_w^2}, \kappa_w+\sqrt{1+\kappa_w^2}, kh_w, \frac{U_s\ssup{0}}{c\ssup{0}}\right) & \text{for } j=w.
\end{cases} \label{eq:alpha_beta_def}
\end{equation}
The amplitude corresponding to a given background shear profile therefore depends on both the wavenumber $k$ and the parameters in the definition of the background profile. 
\par The Gaussian hypergeometric function in \eqref{eq:double-exp-sol} is given by 
\begin{equation}
    F(y_1, y_2, y_3; \xi)\defeq 1 + \frac{y_1y_2}{y_3}\frac{\xi}{1!} + \frac{y_1(y_1+1)y_2(y_2+1)}{y_3(y_3+1)}\frac{\xi^2}{2!}+...
\end{equation}
\par In the case of this double exponential profile \eqref{eq:double-exponential} (Figure \ref{fig:1}), $\varphi_c, \varphi_s$,  $\varphi'(0^+)$, and $\varphi''(0^+)$ may be computed exactly using \eqref{eq:double-exp-sol}, the differentiation identity for hypergeometric functions, and the Rayleigh equation, respectively. In addition if $U_s\ssup{0}=0$, as in the original case considered by \cite{miles1957}, \eqref{eq:c2-sol} may be explicitly, exactly computed, whereas more complex background profiles may necessitate approximation via numerical methods. Assuming $\gamma=0$ (\eqref{def:normalized-parameters}, no surface tension) also as in the original work, the growth rate \eqref{def:linear-growth-rate} is evaluated as
\begin{equation}\label{eq:linear-growth-double-exp}
    \omega\ssup{0}(k) =\eps_\rho k \text{Im }c\ssup{0}_1=-\eps_\rho\left(\frac{g}{k}\right)^{\frac{1}{2}}\frac{\pi}{2} \frac{\frac{d^2}{db^2}U_c\ssup{0}}{\Big|\frac{d}{db} U_c\ssup{0}\Big|}\frac{|\varphi_c|^2}{|\varphi_s|^2}.
\end{equation}
\par In total, the modified growth rate is given by \eqref{eq:linear-growth-double-exp} and \eqref{eq:c2-sol} via
\begin{equation}\label{def:modified-growth-rate}
\omega=  k(\eps_\rho \text{Im }c\ssup{0}_1 + \eps^2\text{Im }c\ssup{2}).
\end{equation}
which is plotted for different values of $\eps$ in Figure \ref{fig:3}. Note that the $\eps_\rho$ and the subscript one are due to the asymptotic expansion done to find $c\ssup{0}$, which is unnecessary when computing $c\ssup{2}$. The formula for $c\ssup{2}$ \eqref{eq:c2-sol} also incorporates the relative densities. Consistent with linear theory (Fig. \ref{fig:3}, solid curves), the growth rate peaks at an intermediate wavenumber and increases with background shear (modified via increasing $U_\infty^{(0)}$). For a finite but nonzero wave slope $\eps$, this growth is systematically suppressed, an effect that intensifies with increasing $\eps$. For moderate values of $U_\infty^{(0)}$, this suppression is largely confined to high wavenumbers, but at larger $U_\infty^{(0)}$ values, it extends even to intermediate wavenumbers, reducing the peak growth rate by as much as a factor of four for the largest wave slope. The aggregate modulation, obtained by trapezoidal integration of the growth curves shown in Figure \ref{fig:3}, exhibits a maximum growth rate decrease (among the parameters shown) of $39.85\%$ for the $\eps=0.3$, $U_\infty=10$ case and a minimum decrease of $0.30\%$ for the $\eps=0.1$, $U_\infty=5.0$ case. 
\begin{figure}[t]
    \centering
    \includegraphics[width=0.8\linewidth]{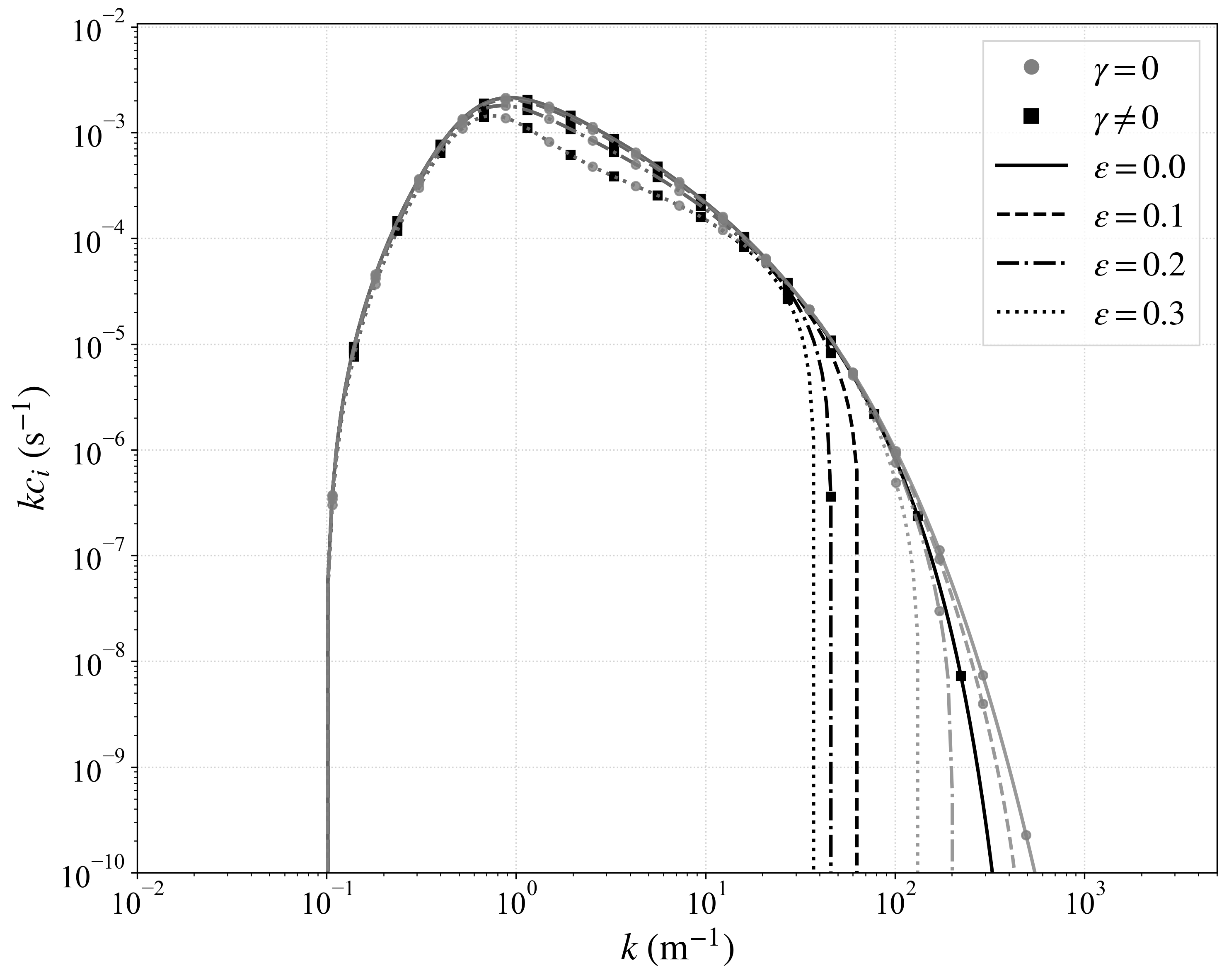}
    \caption{The modified growth rate with capillarity as a function of wavenumber $k$, for four wave slopes $\eps$. The grey curves are as in Figure \ref{fig:3} ($\gamma=0$) whereas the black curves show the impact of including surface tension effects ($\gamma\neq 0$, here $\gamma=7.2\times 10^{-5} \;\text{m}^3\;\text{s}^{-2}$). The parameters are as in Figure \ref{fig:3} but all growth rate curves are with respect to $U_\infty\ssup{0}=10.0 $ m s$^{-1}$. In addition to the modification from the wave-induced current, surface tension further suppresses growth at high wavenumbers. However, surface tension does not alter the effect of the modification to growth rate shown in Figure \ref{fig:3} at intermediate wavenumbers.}
    \label{fig:4}
\end{figure}
\par Up to this point, we have considered the same system as \cite{miles1957} and neglected the effects of surface tension. Since the predicted impacts of the higher-order theory are largest for shorter waves (Figure \ref{fig:3}), which are likely impacted by surface tension, we also compute the impacts of surface tension on both $c\ssup{0}$ and $c\ssup{2}$. The inclusion of surface tension does not change that $c\ssup{1}=0$. The full calculation of these effects involves computing the curvature at points on the surface in Lagrangian coordinates by considering it as a parametric curve $\bm{r}(a)=(x(a;b=0,t),z(a;b=0,t))$ and expanding asymptotically, and is included in Appendix \ref{appC}. A typical value of the kinematic surface tension $\gamma$ is $7.2\times 10^{-5} \;\text{m}^3\;\text{s}^{-2}$ and the modified growth rate computed with this value is shown in Figure \ref{fig:4}. While the intermediate wavenumbers affected by the modified growth rate (\emph{i.e.}, the inclusion of the wave-induced current) have the same growth whether surface tension is included or not, the highest wavenumbers ($k>30\; \text{m}^{-1}$) are significantly impacted by capillary effects. The impacts of surface tension, like those of the inclusion of the wave-induced mean flow, increase with wave slope. In all, the stabilizing effect of surface tension for short waves is amplified by the nonlinear modification of the growth rate shown in Figure \ref{fig:3}.

\section{Discussion and conclusions}
\par Our Lagrangian analysis of the Miles instability reveals the key role Lagrangian drift, including the wave-induced mean flow, plays in modifying the growth of wind-generated waves. While the fundamental instability has long been understood to result from a resonance mechanism, the full role of the wave-induced current remained poorly understood. Our analysis of an idealized but realistic wind profile demonstrates how the growth rate modification, which depends on the wave-induced mean flow, can lead to a suppression of growth with increasing wave steepness. This finding points to an intrinsic self-regulatory mechanism governing the nonlinear evolution of the instability. 
\par To physically interpret this mechanism, we formulate a momentum budget. This momentum budget, expressed in the unapproximated Lagrangian coordinates \eqref{def:monochromatic}, emphasizes that the critical layer interaction is not governed by the Eulerian background shear alone, but by the coupling between the \emph{total} phase speed \emph{c}
 and the \emph{total} Lagrangian mean flow $U$:
 \begin{equation}\label{eq:momentum-budget}
    \frac{\partial}{\partial t}\int_0^{b_c}\mathcal{J}\dot{x} \; \text{d}b + \frac{\partial}{\partial a}\int_0^{b_c} pz_b \, \text{d}b = \rho_0 g z \frac{\dot{z}}{c-U}\Bigg\vert_{b=0}^{b=b_c} - \frac{1}{2}\rho_0(\dot{x}^2+\dot{z}^2)\dot{z}\Bigg\vert_{b=0}^{b=b_c}+\rho_0 f(b,t)\Big\vert^{b=b_c}_{b=0}.
\end{equation}
In \eqref{eq:momentum-budget}, $f(b,t)$ is a constant of integration determined by the phase-averaged pressure at the surface. The momentum budget resulted from integrating the momentum equation, and using an alternative form of pressure which can be derived from the momentum equations (see Appendix \ref{appE}),
\begin{equation}\label{eq:pressure-alternate}
    p = \rho_0 \left(- g z + \frac{1}{2}(c-U)(\dot{x}^2+\dot{z}^2)+f(b,t)\right).
\end{equation}
The first term on the left-hand side of \eqref{eq:momentum-budget} represents the change in momentum density over time, while the second term on the left-hand side is momentum flux. Physically, the second left-hand term accounts for redistribution of momentum laterally within the layer between the surface and the critical level. The right-hand side accounts for the net momentum input across the boundaries at the surface ($b=0$) and the critical level computed in $\S$\ref{sec:2} $(b=b_c)$, which is the physical manifestation of form drag. This form drag is split into two components, originating from the hydrostatic and non-hydrostatic components of the pressure \eqref{eq:pressure-alternate}, respectively. The strength of the forcing associated with the first component is proportional to the wave amplitude via $z(b)$; however, this is modulated by the key phase shift between wave motion and the pressure field. The second component of the form drag takes the form of a vertical flux of kinetic energy, representing the work done by that pressure component at the boundaries. 
\par Our finding that wave growth can be suppressed by its own induced mean flow is compatible with and offers a new perspective on prior observations of wave growth from laboratory and field studies. The suppression of wave growth with increasing wave slope, reflected in both the modified growth rate (Figure \ref{fig:3}) and the momentum budget \eqref{eq:momentum-budget}, is a direct consequence of the wave's nonlinear feedback on its own growth. The dependence of the growth rate on wave steepness emerges from the asymptotic expansion; the well-known $O(\eps^2)$ scaling of the wave-induced mean flow results in a growth rate modification of the same order, providing a causal link between steepness and the suppression. While at linear order the critical layer (defined as $b_c: c\ssup{0}(b_c)=U\ssup{0}(b_c)$, as it is not necessarily the case that there is any $b$ such that $c(b)=U(b)$) resonance is perfect and the momentum transfer is maximally efficient, the higher-order induced mean flow creates a mismatch between the total phase speed and Lagrangian mean flow ($c\neq U$ in general). This detunes the resonance and ultimately results in less efficient momentum transfer. Note that this mean-flow mechanism is distinct from a simple alteration of the surface current (as considered in prior studies); it modifies the Lagrangian flow profile without changing the overall shear of the background Eulerian flow. Thus, while the observed growth suppression \citep{peirson-garcia2008} has been explained as a shift from a highly efficient, extrinsic ``maser-like" mechanism \citep{longuethiggins1969} dominant at low steepness, our analysis shows it can be understood as an inherent consequence of the wave's own dynamics. 
\par By remaining entirely within the Lagrangian frame, our formulation of the momentum budget \eqref{eq:momentum-budget} most faithfully captures Lighthill's \citeyear{Lighthill1962} parcel-based argument, wherein displaced fluid parcels create an asymmetric pressure field that exerts a net force on the wave. The Lagrangian perspective highlights the viewpoint that the instability results from a coupling between different wavy fluid layers: the surface wave itself ($b=0$) and the wave-like motion of fluid parcels within the shear flow (at each level in $b$). At the critical layer, leading-order resonance makes this coupling particularly effective, generating the phase shift between the interior pressure field and the surface elevation. This detailed agreement between the mathematical formalism and the physical intuition underscores the power of the Lagrangian approach for analyzing shear instabilities. Indeed, even the particle trajectories themselves (Figure \ref{fig:1}) provide a direct visual for the instability mechanism: purely circular orbits, corresponding to zero Lagrangian mean flow, would never develop the phase shift necessary for instability. 
\par More broadly, this work establishes a framework for analyzing shear instabilities entirely in the Lagrangian frame. Unlike traditional Eulerian analyses that can obscure the particle dynamics central to physical arguments like \cite{Lighthill1962}, the Lagrangian approach makes them central to the analysis. This methodology is well-suited for revealing the impact of wave-induced flow, and can be readily applied to other canonical shear flow problems to gain new physical insights. 
\par The conclusion that wave growth is governed by the Lagrangian drift connects directly to observational work, as this is precisely the quantity measured by remote sensing techniques that invert for upper ocean currents based on Doppler shifting of surface waves \citep{pizzo2023}. Consequently, observational data gathered via these methods already capture the  wave-induced feedback shown here to modify growth. This provides a direct pathway to reconciling long-standing discrepancies between theory and field measurements, by incorporating observational data that measure the full Lagrangian flow. The physical insights enabled by the non-linear stability analysis offer a pathway for refining wind stress parameterizations, particularly for steeper waves or in high-shear conditions where the higher-order modification is most significant. 
\par Our findings emphasize a key point: the wind interacts with the Lagrangian, rather than Eulerian, mean surface flow. This has implications for the concept of ``relative wind," which is central to air-sea interaction. For instance, the ``eddy killing" mechanism, in which wind extracts kinetic energy from mesoscale ocean eddies, is a significant dissipation pathway in the ocean's energy budget \citep{rai2021}. The observed spatial variability of the wave field (\textit{e.g.}, Figure 2 in \citealt{lenain2023airborne}) implies a corresponding variability in this feedback due to the wave-induced component of the Lagrangian drift --- the quantification of which is a topic for future work. A broader implication of this work is therefore that the wave-induced component of the surface flow may be necessary to include when accurately modeling the atmospheric response to ocean dynamics.



\backsection[Funding]{LRS was supported by the Department of Defense (DoD) through the National Defense Science and Engineering Graduate (NDSEG) Fellowship Program. MAF and NP were supported by NASA grant 80NSSC23K0985. }

\backsection[Declaration of interests]{The authors report no conflict of interest.}


\backsection[Author ORCIDs]{L.R. Seitz, \url{https://orcid.org/0009-0004-1933-1931}, Mara A. Freilich, \url{https://orcid.org/0000-0003-0487-8518}, and Nick Pizzo, \url{https://orcid.org/0000-0001-9570-4200}}

\backsection[Author contributions]{LRS: mathematical analysis; theoretical framework development; writing – original draft. MAF: conceptualization; project guidance. NP: conceptualization; theoretical framework development. All authors contributed to reviewing and editing.}

\begin{appen}
\section{Mathematical details of linear stability analysis}\label{appA}
When expanding $x,z$ and $p$ asymptotically as in \eqref{def:monochromatic-asymp-x}-\eqref{def:monochromatic-asymp-p}, we can also expand the Jacobian \eqref{def:jacobian} and the vorticity \eqref{def:vorticity} in the same fashion, so that 
\begin{subequations}
    \begin{align}
        \mathcal{J} &= \mathcal{J}\ssup{0}+\eps\mathcal{J}\ssup{1} + \eps^2 \mathcal{J}\ssup{2} + \eps^3 \mathcal{J}\ssup{3} + O(\eps^4) \\
        \mathcal{\Gamma} &= \mathcal{\Gamma}\ssup{0}+\eps\mathcal{\Gamma}\ssup{1} + \eps^2 \mathcal{\Gamma}\ssup{2} + \eps^3 \mathcal{\Gamma}\ssup{3} + O(\eps^4).
    \end{align}
\end{subequations}
\par To derive the Rayleigh equation, we first find the linear governing equations by substituting \eqref{def:monochromatic-asymp-x}-\eqref{def:monochromatic-asymp-p} into \eqref{eq:euler-in-lagrangian-x}-\eqref{eq:euler-in-lagrangian-z}. Note that this portion of the derivation of the instability growth rate (deriving the Rayleigh equation) is not novel and was done in \cite{bennett2006}. The linearized equations are given by
\begin{subequations}
    \begin{align}
        \ddot{x}\ssup{1} + gz_a\ssup{1} &= -\frac{1}{\rho_0}p_a\ssup{1}\\ 
        \ddot{z}\ssup{1} - gx_a\ssup{1} &= -\frac{1}{\rho_0}\mathcal{D}p\ssup{1} \\
        x_a\ssup{1} + \mathcal{D}z\ssup{1}&=0 \label{eq:linear-mass-continuity}
    \end{align}
\end{subequations}
where 
\begin{equation}\label{def:d-op}
\mathcal{D} \defeq \partial_b - t\frac{d}{db}U\ssup{0}(b)\partial_a.
\end{equation}
Note that \eqref{eq:linear-mass-continuity} is mass continuity; $\mathcal{J}=\mathcal{J}_0$ for all time, and $\mathcal{J}\ssup{1}=0$. Differentiating further, 
\begin{align*}
    \ddot{z}_{aa}\ssup{1} - gx_{aaa}\ssup{1} = -\mathcal{D} ( \partial_{tt}\mathcal{D}z\ssup{1}) + \mathcal{D}gz_{aa}\ssup{1}.
\end{align*}
Since $-gx_{aaa}\ssup{1}-\mathcal{D}gz_{aa}\ssup{1} = -g\partial_{aa}(x_a\ssup{1}+\mathcal{D}z\ssup{1})=0$,  we conclude 
\begin{equation}\label{eq:lin-z-4th-deriv}
\partial_{tt}\partial_{aa}{z}\ssup{1} + \mathcal{D}(\partial_{tt} \mathcal{D}z\ssup{1})=0.
\end{equation}
We now apply the change of variables \eqref{eq:change-of-vars-z}. After using (\ref{eq:lin-z-4th-deriv}) and the chain rule, \textit{i.e.} $\mathcal{D} z\ssup{1} = \partial_b B$ (where this derivative refers to differentiating with respect to the second slot), etc., we obtain exactly the Rayleigh equation (but in Lagrangian coordinates), \eqref{eq:rayleigh-lagrangian}.
\par The Rayleigh equation is used to derive the identity for $\text{Im } \mathcal{D}_1$, which is used when computing the asymptotic approximation \eqref{eq:imc1-2}. The identity \eqref{eq:growth-rate-linear} is analogous to that in the appendix of \cite{young2014}, but is included here for completeness. By multiplying the Rayleigh equation \eqref{eq:rayleigh-lagrangian} by $\varphi^*(b)$, integrating from $b=0$ to $b=\infty$, and using integration by parts, one finds that 
\begin{subequations}
\begin{align}
    \Xi_{air}(c\ssup{0},k) &= \frac{1}{|\varphi_s|^2}\int_0^\infty |\varphi'|^2 + \left(k^2 + \frac{\frac{d^2}{db^2}U\ssup{0}}{U\ssup{0}-c\ssup{0}}\right)|\varphi|^2\;\text{d}b, \label{eq:xi-air-int} \\
    \Xi_w(c\ssup{0},k) &= \frac{1}{|\varphi_s|^2}\int_{-\infty}^0 |\varphi'|^2 + \left(k^2 + \frac{\frac{d^2}{db^2}U\ssup{0}}{U\ssup{0}-c\ssup{0}}\right)|\varphi|^2\;\text{d}b.
\end{align}
\end{subequations}
Only the integrand of \eqref{eq:xi-air-int} has a singularity, which is at $b_c: U\ssup{0}(b_c)=c\ssup{0}$, because we assume $b_c$ is in the air ($b_c>0$). To avoid the unphysical singularity, we consider a \emph{near} singularity, and introduce the complex phase speed \eqref{eq:complex-phase-speed}. We then approach the singularity by taking $c_i\to 0$. We compute that 
\begin{equation}
\text{Im } \left(\frac{\frac{d^2}{db^2} U\ssup{0}}{U\ssup{0}-c_r\ssup{0}-ic_i\ssup{0}}\right) =  \frac{\frac{d^2}{db^2}U\ssup{0}c_i\ssup{0}}{(U\ssup{0}-c_r\ssup{0})^2+(c_i\ssup{0})^2}.
\end{equation}
Then we can use the following identity (which can be seen from the relationship of the pdf of a Cauchy distribution and the delta function, though in \cite{miles1957} this is done using residue calculus):
\begin{equation}\label{eq:im-xiair-identity}
\lim_{c_i\ssup{0} \to 0} \frac{c_i\ssup{0}\frac{d^2}{db^2}U\ssup{0}}{(U\ssup{0}-c_r\ssup{0})^2+(c_i\ssup{0})^2}=\pi\frac{\frac{d^2}{db^2}U\ssup{0}_c}{\Big|\frac{d}{db}U\ssup{0}_c\Big|}\delta(b-b_c).
\end{equation}
The $\frac{d^2}{db^2}U_c\ssup{0}$ in the numerator of \eqref{eq:im-xiair-identity} is because of that extra factor in the numerator, while the denominator $|\frac{d}{db}U\ssup{0}(b_c)|$ arises from the change of variables formula for the Dirac delta. 

\section{Mathematical details of nonlinear stability analysis}\label{appB}
The dispersion relation \eqref{eq:c2-disp-relation-no-gamma} that describes the role of the wave-induced mean flow on the instability growth rate results from considering the system at third order in the wave slope. We first find the dispersion relation at both linear order \eqref{eq:disp-relation-lin-abst-1} and order $\eps^2$. We will consider the case originally considered by \cite{miles1957}, and neglect the effects of surface tension, setting $\gamma=0$. The task then reduces to tabulating the pressure jump at each order, which requires computing each term in \eqref{eq:pressure}. Obtaining informative results requires significant simplification of this expression, which requires enforcing the conservation of mass and vorticity at each order.
\par Since we must expand $c=c\ssup{0}+\eps c\ssup{1} + \eps^2 c\ssup{2} + O(\eps^3)$, we rewrite the expansions of $x$ \eqref{def:monochromatic-asymp-x} and $z$ \eqref{def:monochromatic-asymp-z} in terms of wave slope $\eps$ in more detail. Through second order,
\begin{equation}
k(a-(c-U(b))t) = k(a-(c\ssup{0} + \eps c\ssup{1} +\eps^2c\ssup{2} - U\ssup{0}(b) -\eps^2U\ssup{2}(b))t).
\end{equation}
We will not \emph{a priori} expand $U\ssup{2}$ or $U\ssup{3}$ and we will neglect the possible dependence of $x_n(b), z_n(b)$ on $c\ssup{2}, c\ssup{3}$ for now, though this will later be accounted for when we explicitly apply a generalization of the change of variables \eqref{eq:change-of-vars-x}-\eqref{eq:change-of-vars-z}. Otherwise expanding,
\begin{subequations}
\begin{align}
\eps x_1(b)&\sin(k(a-(c^{(0)} + \eps c\ssup{1}+\eps^2c^{(2)} - U^{(0)}(b) -\eps^2U^{(2)}(b))t)) \approx 
\eps x_1(b) \sin(\theta_0) 
- \eps^2  x_1(b) ktc^{(1)}\cos(\theta_0) \nonumber\\
&- \eps^3  x_1(b) \frac{(ktc^{(1)})^2}{2}\sin(\theta_0) 
-\eps^3 x_1(b) kt(c^{(2)} - U^{(2)}(b))\cos(\theta_0) + O(\eps^4) \label{eq:full-expansion-x} \\
\eps z_1(b) &\cos(k(a-(c^{(0)} + \eps c\ssup{1}+\eps^2c^{(2)} - U^{(0)}(b) -\eps^2U^{(2)}(b))t)) \approx 
\eps  z_1(b) \cos(\theta_0) + \eps^2  z_1(b) ktc^{(1)}\sin(\theta_0) \nonumber \\
&-\eps^3 z_1(b) \frac{(ktc\ssup{1})^2}{2}\cos(\theta_0) + \eps^3  z_1(b)  kt(c^{(2)} - U^{(2)}(b))\sin(\theta_0)+ O(\eps^4) \label{eq:full-expansion-z}
\end{align}
\end{subequations}
Here, $\theta_0 \defeq \thetzero$, for ease of notation. Besides one more $\eps^3$ contribution from the second order terms (\textit{i.e.} $\sin2\thet , \cos 2\thet )$ due to $c\ssup{1}$, the higher order oscillatory terms will not provide any additional contributions through third order. We do not expand further because at second order, we will see $c\ssup{1}=0$ is required for conservation of vorticity. 
\par Starting at order $\eps$, the conservation of mass condition, $\dot{\mathcal{J}}\ssup{1}=0$, simplifies to 
\begin{equation}\label{eq:mass-cont-oeps}
    kx_1(b) = -z_1'(b),
\end{equation}
where the use of primes throughout this section will always denote differentiation with respect to $b$. The condition \eqref{eq:mass-cont-oeps} implies $\mathcal{J}\ssup{1}=0$. 
It can also be computed that $\mathcal{J}\ssup{0}=1$. 
\par First order conservation of vorticity requires 
\begin{equation}\label{eq:cons-vort-linear}
   (c^{(0)}-U\ssup{0}) (k^{2} z_{1}{\left(b \right)} - z_1''(b)) + 2 \frac{d}{d b} U^{(0)}{\left(b \right)} z_1'(b)=0,
\end{equation}
which implies $\Gamma\ssup{0} =0$. Using these facts, we can evaluate each term in \eqref{eq:pressure} to obtain the pressure-jump portion of dispersion relation \eqref{eq:disp-relation-changed-var-lin-fin}, after changing variables. The linear pressure is evaluated as 
\begin{equation}
    p\ssup{1}(a,b,t) = -\rho_0\Big(gz_1+ k (c\ssup{0}-U\ssup{0})^2 x_{1}{\left(b \right)}\Big) \cos(\theta_0).
\end{equation}

\par Conservation of mass at second order requires that 
\begin{equation}\label{eq:mass-cons-2nd-order}
(2kx_2+z_2')-\frac{1}{2}(z_1z_1''-(z_1')^2)=0,
\end{equation}
which implies the Jacobian is given by 
\begin{equation}
    \mathcal{J}\ssup{2} = -\frac{1}{2}(z_1z_1''+(z_1')^2).
\end{equation}
Conservation of vorticity at second order requires
\begin{equation}\label{eq:cons-vort-oeps2}
2(c\ssup{0}-U\ssup{0})(2k^2z_2+kx_2')-\left(4kx_2+\frac{1}{2}k^2(x_1^2-z_1^2)\right)\frac{d}{db}U\ssup{0}=0
\end{equation}
and importantly that  
\begin{equation}\label{eq:cons-vort-oeps2-2}
c\ssup{1}=0.
\end{equation}
The condition \eqref{eq:cons-vort-oeps2-2} is necessary to avoid time-dependence of the vorticity, but also eliminates secular growth. Then the vorticity term (\emph{i.e.} vorticity multiplied by the Jacobian) is simply 
\begin{equation}\label{eq:vort-second-order}
    (\Gamma\mathcal{J})\ssup{2}= \frac{1}{2}\frac{d}{db}\Big((c^{(0)}-U^{(0)})k^2(x_1^2+z_1^2)\Big) - \frac{d}{db}U^{(2)}.
\end{equation}
Assembling the contributions to the pressure, we find that 
\begin{equation}\label{eq:pressure-oeps2}
\begin{split}
p\ssup{2}(a,b,t) &= -g\left(z_0\ssup{2}+z_2(b)\cos(2\theta_0)\right)
-\frac{1}{2}\left((c^{(0)} - U^{(0)})^2\left(\frac{1}{2}k^2(x_1^2+z_1^2) + 4k x_2(b)\right)\right) \cos(2\theta_0) \\
&\quad+ \frac{1}{2}\int_{-\infty}^b (c\ssup{0}-U\ssup{0})^2k^2(x_1x_1'+z_1z_1')\;\text{d}\beta.
\end{split}
\end{equation}
However, \eqref{eq:pressure-oeps2} cannot be used to yield a dispersion relation reflecting the impact of the wave-induced mean flow, so we continue to third order. 
 \par Conservation of mass at third order requires
\begin{subequations}
    \begin{align}
        kx_2z_1' + \frac{k}{2}x_2'z_1- z_2z_1''-\frac{1}{2}z_1'z_2'&=0 \label{eq:cons-mass-eps3-1} \\
3kx_3+ z_3'+kx_2z_1' - \frac{k}{2}x_2'z_1+z_2z_1'' - \frac{1}{2}z_1'z_2' &=0 \label{eq:cons-mass-eps3-2}
    \end{align}
\end{subequations}
and the Jacobian at this order is $\mathcal{J}\ssup{3}=0$. 
\par Conservation of vorticity at third order requires that 
\begin{subequations}
\begin{align}
    -2k^2(x_1x_2+z_1z_2)\frac{d}{db}U\ssup{0}-z_1\frac{d^2}{db^2}U\ssup{2} + \frac{c\ssup{2}-U\ssup{2}}{c\ssup{0}-U\ssup{0}}z_1\frac{d^2}{db^2} U\ssup{0} &\nonumber\\
    +\frac{3}{2}k^2(c\ssup{0}-U\ssup{0})(x_1x_2'+2x_2x_1'+z_1z_2'+2z_2z_1')&=0 \label{eq:cons-vort-oeps3-1}\\ 
    3(c^{(0)}-U\ssup{0})(3k^{2} z_3 + k x_3')- 6 k x_3 \frac{d}{d b} U^{(0)}-2k^2\frac{d}{db}U\ssup{0}(x_1x_2-z_1z_2)& \nonumber\\
    +\frac{1}{2}k^2(c\ssup{0}-U\ssup{0})(x_1x_2'+2x_2x_1'+z_1z_2'+2z_2z_1')&=0 \label{eq:cons-vort-oeps3-2}
\end{align}
\end{subequations}
To this point though, we have not considered that we want to make a change of variables such that 
\begin{equation}\label{eq:z1-dependence-on-c2}
z_1 = \frac{\varphi}{k(c\ssup{0}-U\ssup{0} + \eps^2( c\ssup{2} -U\ssup{2}))}.
\end{equation}
We would make an analogous change of variables for $z_2, z_3,$ etc. as well, but the impact of $c\ssup{2}$  on these terms (and the impact of $c\ssup{n+2}$ on $z_n$, $n\geq 1$) only comes in at $O(\eps^4)$ so we do not explicitly consider those contributions here. This is also why we omit $\eps^3(c\ssup{3}-U\ssup{3})+O(\eps^4)$ from the denominator of \eqref{eq:z1-dependence-on-c2}. Note that we neglect any possible dependence of $\varphi$ on $c\ssup{n}, n\geq 2$ when finding additional terms, as we do not assume a specific form of $\varphi$ \emph{a priori}. If the modified growth rate we calculate is applied to wind profiles that yield $\varphi$ that do depend on $c$, this amounts to assuming the dominant impact of $c$ in the amplitude is through its appearance in the denominator from the change of variables. Given a particular definition of $\varphi$, one could also account for the impacts of this term \emph{a posteriori} in \eqref{eq:appendix-third-order-disp}, as the term will not be secular.  When we expand using \eqref{eq:z1-dependence-on-c2}, we obtain
\begin{equation}
z_1=\frac{\varphi}{k(c\ssup{0}-U\ssup{0})}+\eps^3 \tilde{z}_1\cos\thetzero + O(\eps^4).
\end{equation}
Solving for $\tilde{z}_1$, the additional term in the equation for $z$  \eqref{def:monochromatic-asymp-z} is then given by
\begin{equation}\label{eq:third-order-bonus-term-z}
 \eps^3 \tilde{z}_1\cos\thetzero \text{ where } \tilde{z}_1 = -\frac{c\ssup{2}-U\ssup{2}}{c\ssup{0}-U\ssup{0}}z_1(b).
\end{equation}
In the equation for $x$, we obtain another term as well,
\begin{equation}\label{eq:third-order-bonus-term-x-0}
\eps^3\tilde{x}_1\sin\thetzero,
\end{equation}
where \eqref{eq:mass-cont-oeps} implies 
\begin{equation}\label{eq:third-order-bonus-term-x}
\tilde{x}_1 = \frac{1}{k}\left(\frac{c\ssup{2}-U\ssup{2}}{c\ssup{0}-U\ssup{0}}\right)z_1' +\frac{1}{k}\frac{d}{db}\left( \frac{c\ssup{2}-U\ssup{2}}{c\ssup{0}-U\ssup{0}}\right)z_1.
\end{equation}
The extra terms \eqref{eq:third-order-bonus-term-z} and \eqref{eq:third-order-bonus-term-x} add zero to the Jacobian, leaving \eqref{eq:cons-mass-eps3-1} and \eqref{eq:cons-mass-eps3-2} unchanged. 
\par At third order, the kinetic energy term in \eqref{eq:pressure} is complicated, so we write out the evaluation in more detail here. The kinetic energy term, first without the additional terms due to the dependence of $z_1$ on $c\ssup{2}$ (\emph{i.e.} \eqref{eq:z1-dependence-on-c2}), is: 
\begin{equation}\label{eq:ke-oeps3}
\begin{split}
    -&\frac{(\dot{x}-c)^2 + \dot{z}^2}{2}\Bigg|_{\eps^3} = -(c^{(0)}-U\ssup{0})^{2}\left((c^{(2)}-U\ssup{2}) k^{2} t x_1\sin(\theta_0) +3  k x_3\cos(3\theta_0) \right)- (c\ssup{0}-U^{(0)})(c\ssup{3}- U^{(3)})\\
    &-2 (c^{(0)}-U\ssup{0}) \left( (c^{(0)}-U\ssup{0})k^{2} (x_1x_2\cos(\theta_0) \cos(2\theta_0) + z_1z_2\sin(\theta_0) \sin(2\theta_0)) + ( c^{(2)} - U\ssup{2})k x_1 \cos(\theta_0) \right)
\end{split}
\end{equation}

\par Before assembling the contributions to the dispersion relation at this order, recall from first order
\begin{equation}\label{eq:linear-dispersion-in-z}
    p^{(1)}(a,0^+,t)-p^{(1)}(a,0^-,t)= (-\rho_{air}gz_1(0^+)+\rho_{w}gz_1(0^-)+(c\ssup{0}-U_s\ssup{0})^2k(\rho_{w}x_1(0^-)-\rho_{air}x_1(0^+)))\cos(\theta_0).
\end{equation}
Since we neglect the effects of surface tension, we intend to set the pressure jump equal to zero. We then see that the secular terms in kinetic energy expression \eqref{eq:ke-oeps3}, together with the secular term that would come from $gz$, \emph{i.e.} as shown in \eqref{eq:full-expansion-z}, will contribute zero to the third order dispersion relation. This is because \eqref{eq:linear-dispersion-in-z} in the case of no surface tension implies 
\begin{equation}\label{eq:linear-dispersion-in-z-no-surface-tension}
-\rho_{air}(gz_1(0^+)+(c\ssup{0}-U_s\ssup{0})^2kx_1(0^+))-\rho_{w}(gz_1(0^+)+(c\ssup{0}-U_s\ssup{0})^2kx_1(0^+))=0
\end{equation}
which would still hold when multiplied by, \emph{e.g.}, $(c\ssup{2}-U\ssup{2}_s)kt\sin(\theta_0)$.
\par When considering the terms arising from the dependence of $z_1$ on $c\ssup{2}$ and computing the kinetic energy and the gravitational term, the contribution of $\tilde{z}_1$ and the first term in $\tilde{x}_1$ will reduce to zero due to the linear dispersion relation. However, the last term in $\tilde{x}_1$ \eqref{eq:third-order-bonus-term-x} remains, and it can be seen that its contribution via the kinetic energy is 
\begin{equation}
z_1\left((c\ssup{0}-U\ssup{0})\frac{d}{d b}U\ssup{2}-(c\ssup{2}-U\ssup{2})\frac{d}{d b}U\ssup{0}\right)\cos(\theta_0).
\end{equation}
The remaining contributions to the third order dispersion relation are (omitting those whose contributions will cancel out, as discussed previously):
\begin{align} 
-gz|_{\eps^3} &= -gz_3(b)\cos (3\theta_0) - gz_0\ssup{3} \\
-\frac{(\dot{x}-c)^2 + \dot{z}^2}{2}\Bigg|_{\eps^3}&= -2k^2(c\ssup{0}-U\ssup{0})^2(x_1x_2\cos(\theta_0)\cos(2\theta_0)+z_1z_2\sin(\theta_0)\sin(2\theta_0)) \nonumber\\
+\Big(kz_1&\Big((c\ssup{0}-U\ssup{0})\frac{d}{d b}U\ssup{2}-(c\ssup{2}-U\ssup{2})\frac{d}{d b}U\ssup{0}\Big)- 2kx_1(c^{(0)} - U^{(0)})(c^{(2)} - U^{(2)})\Big) \cos(\theta_0) \nonumber\\
&- 3k(c^{(0)} - U^{(0)})^2x_3\cos(3\theta_0) - (c^{(0)} - U^{(0)})(c\ssup{3}-U^{(3)}) \\
\int_{-\infty}^b(c-U)\Gamma\mathcal{J} \,\text{d}\beta \Bigg|_{\eps^3}&= -\int_{-\infty}^b (c\ssup{0}-U\ssup{0}(\beta))\frac{d}{d\beta}U\ssup{3}(\beta)\,\text{d}\beta-\int_{-\infty}^b (c\ssup{3}-U\ssup{3}(\beta))\frac{d}{d\beta}U\ssup{0}(\beta) \,\text{d}\beta \nonumber\\
&= c\ssup{0}U\ssup{3}-c\ssup{3}U\ssup{0}+U\ssup{0}U\ssup{3} \\
f(a,t)\Big|_{\eps^3} &= c\ssup{0}c\ssup{3}
\end{align}
The pressure is then
\begin{equation}\label{eq:pressure-eps3-1}
\begin{split}
p\ssup{3} &= -\rho_0(gz_3+3k(c\ssup{0}-U\ssup{0})x_3)\cos(3\theta_0) - g\rho_0z_0\ssup{3} \\
&- \rho_0\left(2k(c\ssup{0}-U\ssup{0})(c\ssup{2}-U\ssup{2})x_1+z_1\left((c\ssup{0}-U\ssup{0})\frac{\partial}{\partial b}U\ssup{2}-(c\ssup{2}-U\ssup{2})\frac{\partial}{\partial b}U\ssup{0}\right)\right)\cos(\theta_0) \\
&-2\rho_0(c\ssup{0}-U\ssup{0})^2k^2(x_1x_2\cos(\theta_0)\cos(2\theta_0)+z_1z_2\sin(\theta_0)\sin(2\theta_0)).
\end{split}
\end{equation}
We can rewrite \eqref{eq:pressure-eps3-1} using the trigonometric identities 
\begin{align}
    \cos(\theta_0)\cos(2\theta_0) &
= \frac{1}{2}(\cos(-\theta_0) + \cos(3\theta_0)) = \frac{1}{2}(\cos(\theta_0) + \cos(3\theta_0)) \\
\sin(\theta_0)\sin(2\theta_0) &= \frac{1}{2}(\cos(-\theta_0)-\cos(3\theta_0))=\frac{1}{2}(\cos(\theta_0)-\cos(3\theta_0))
\end{align}
as
\begin{equation}\label{eq:pressure-eps3-2}
\begin{split}
p\ssup{3} &= -\rho_0(gz_3+(3k(c\ssup{0}-U\ssup{0})x_3+(c\ssup{0}-U\ssup{0})^2k^2(x_1x_2-z_1z_2)))\cos(3\theta_0 )- \rho_0gz_0\ssup{3} \\
&- \rho_0\Big(2k(c\ssup{0}-U\ssup{0})\Big((c\ssup{2}-U\ssup{2})x_1- (c\ssup{0}-U\ssup{0})k(x_1x_2+z_1z_2)\Big) \\
&\quad+z_1\left((c\ssup{0}-U\ssup{0})\frac{d}{db}U\ssup{2} - (c\ssup{2}-U\ssup{2})\frac{d}{db}U\ssup{0}\right) \Big)\cos(\theta_0). 
\end{split}
\end{equation}
Since $\cos(\theta_0)$ and $\cos(3\theta_0)$ (and a constant) are linearly independent functions, we can satisfy the pressure-jump condition by equating the coefficient of each term to zero independently. Doing so for the $\cos(\theta_0)$ term reveals that $c\ssup{2}$ satisfies the dispersion relation 
\begin{equation}\label{eq:appendix-third-order-disp}
\begin{split}
    \rho_{air}&\Bigg(-2z_1'(0^+)(c\ssup{0}-U\ssup{0}_s)(c\ssup{2}-U\ssup{2}(0^+))-2(c\ssup{0}-U_s\ssup{0})^2(-kz_1'(0^+)x_2(0^+)+k^2z_1(0^+)z_2(0^+))\Bigg)\\
    & +z_1(0)\Bigg((c\ssup{0}-U_s\ssup{0})\frac{d}{db}U\ssup{2}(0^+) - (c\ssup{2}-U\ssup{2}(0^+))\frac{d}{db}U\ssup{0}(0^+)\Bigg)\\
    -\rho_{w}&\Bigg(-2z_1'(0^-)(c\ssup{0}-U\ssup{0}_s)(c\ssup{2}-U\ssup{2}(0^-))-2(c\ssup{0}-U_s\ssup{0})^2(-kz_1'(0^-)x_2(0^-)+k^2z_1(0^-)z_2(0^-)) \\
&+ z_1(0)\Bigg((c\ssup{0}-U_s\ssup{0})\frac{d}{db}U\ssup{2}(0^-) - (c\ssup{2}-U\ssup{2}(0^-))\frac{d}{db}U\ssup{0}(0^-)\Bigg)\Bigg)=0,
\end{split}
\end{equation}
as in $\S$\ref{subsec:nonlinear-stab-analysis}.
\section{Inclusion of surface tension effects}\label{appC}
Surface tension effects appear on the right-hand side of the dynamic boundary condition, \emph{e.g.} in \eqref{eq:linearized-dynamic-boundary}. The curvature effects are given by $T\kappa$ where $T$ is the coefficient of surface tension and $\kappa$ is the curvature at a given point on the water surface. In Lagrangian coordinates, for a fixed time $t$, the surface is described by the parametric curve $\bm{r}(a)=(x(a;b=0,t),z(a;b=0,t))$. The curvature for a parametric vector is given by (using the convention of signed curvature)
\begin{equation}\label{def:curvature-param}
    \kappa = \frac{\bm{r}_a\times \bm{r}_{aa}}{|\bm{r}_a|^3}.
\end{equation}
We compute
\begin{subequations}
\begin{align}
|\bm{r}_a\times \bm{r}_{aa}| &= |x_az_{aa}-z_ax_{aa}|, \label{eq:surface-curv-calc-1} \\
|\bm{r}_a|^3 &= (x_a^2+z_a^2)^{\frac{3}{2}}. \label{eq:surface-curv-calc-2}
\end{align}
\end{subequations}
Upon substituting \eqref{eq:surface-curv-calc-1} and \eqref{eq:surface-curv-calc-2} into \eqref{def:curvature-param}, we obtain
\begin{equation}\label{eq:curvature-lag-coords}
\kappa = \frac{x_az_{aa}-z_ax_{aa}}{(x_a^2+z_a^2)^{3/2}}.
\end{equation}
Note that \eqref{eq:curvature-lag-coords} is also obtained upon transforming the Eulerian definition of curvature, with respect to the free surface $z=\eta(x)$,
\begin{equation}
    \kappa = \frac{\eta_{xx}}{(1+\eta_x^2)^{3/2}}
\end{equation}
into Lagrangian coordinates directly. We now evaluate \eqref{eq:curvature-lag-coords} at each order of $\eps$, using \eqref{eq:full-expansion-x}, \eqref{eq:full-expansion-z}, including the additional terms due to the change of variables, $\tilde{x}_1$ \eqref{eq:third-order-bonus-term-x} and $\tilde{z}_1$ \eqref{eq:third-order-bonus-term-z}. We obtain, again denoting $\theta_0\defeq \thetzero$:
\begin{align}
    T\kappa|_\eps &= -Tk^2z_1(0)\cos(\theta_0),\\
    T\kappa|_{\eps^2} &=  Tk^2(-kx_1(0)z_1(0)\sin^2(\theta_0)+2kx_1(0)z_1(0)\cos^2(\theta_0)-4z_2(0)\cos(2\theta_0))\\
        T\kappa|_{\eps^3} &=-(c\ssup{2}-U\ssup{2}_s) k^{3} t z_1(0)\sin(\theta_0)+ \left(-k^{2}  \tilde{z}_1(0)- \frac{3}{2} k^{4} x_1^2(0) z_1(0) + \frac{3}{8} k^{4} z_1^3(0) + 3 k^{3} x_1(0) z_2(0)\right) \cos(\theta_0), \\
&+ \left(- \frac{3}{2} k^4 x_1^2(0)z_1(0) - \frac{3}{8} k^4 z_1^3(0)+ 5 k^3 x_1(0) z_2(0)+ 4 k^{3} z_1(0)x_2(0)- 9 k^2z_3(0)\right) \cos(3\theta_0).
\end{align}
Notice that the secular term and the term with $\tilde{z}_1$ are exactly those needed to generalize \eqref{eq:linear-dispersion-in-z-no-surface-tension} so that the contributions from the secular term and $\tilde{z}_1$ in the dispersion relation still reduce to zero. Then, again due to the linear independence of $\cos(\theta_0)$ and $\cos(3\theta_0)$, the third-order surface tension effects only contribute an extra term of 
\begin{equation}
    \left( - \frac{3}{2} k^{4} x_{1}^2{\left(0 \right)} z_{1}{\left(0 \right)} + \frac{3}{8} k^{4} z_{1}^3{\left(0 \right)} + 3 k^{3} x_{1}{\left(0 \right)} z_{2}{\left(0 \right)}\right)
\end{equation}
to the right-hand side of \eqref{eq:appendix-third-order-disp}, which consequently remains linear in $c\ssup{2}$. Note that the matter of continuity of $z_2$ at zero is irrelevant to the case $U_s\ssup{0}=0$ but may require more careful consideration for more complex background shear profiles. Additionally, the inclusion of surface tension effects yields the solution $c\ssup{0} = \left(\frac{g}{k} + \gamma k\right)^{\frac{1}{2}}$ per \eqref{eq:leading-order-balance-disp-lin}.
\section{Explicit formula for the wave-induced mean flow}\label{appD}
We derive the form of the (second-order) wave-induced mean flow $U\ssup{2}$ -- or otherwise stated, the Lagrangian frame analogue of Stokes drift -- without making any assumptions on vorticity, by adapting Stokes' \citeyearpar{stokes1847} original argument to Lagrangian coordinates and to the situation in which there is a leading-order Lagrangian mean flow. 
\par Let $\xi$ denote the deviation of a particle's $x$-position from the background state $a+U\ssup{0}(b)t$ and $\eta$ denote the deviation of a particle's $z$-position from the background state $b$. We will find the values of $\xi$ and $\eta$ occurring at second order in the wave slope by using the forms of $x\ssup{1}$ and $z\ssup{1}$. We compute to first order (omitting order $\eps^0$ terms since they already appear in the series expansions for $x$ \eqref{def:monochromatic-asymp-x} and $z$ \eqref{def:monochromatic-asymp-z})
\begin{equation}\label{eq:xi-approx-1}
    \frac{d\xi}{dt} = u \approx -\eps kx_1(b+\eta)(c\ssup{0}-U\ssup{0}(b+\eta))\cos(k(a-\xi-(c\ssup{0}-U\ssup{0}(b+\eta))t)). 
\end{equation}
Inside of the cosine, we approximate $U\ssup{0}(b+\eta)\approx U\ssup{0}(b)$ since we are only concerned with order $\eps^2$ and lower terms. In this section, we again denote $\theta_0\defeq \thetzero$ for compactness. Then \eqref{eq:xi-approx-1} becomes
\begin{equation}
\begin{split}
\frac{d\xi}{dt} &\approx -\eps k(x_1(b)+x_1'(b)\eta)(c\ssup{0}-U\ssup{0}(b))(\cos(\theta_0)-k\xi\sin(\theta_0)) \\
    &= \eps(c\ssup{0}-U\ssup{0}(b))\left( \xi k^2 x_1(b) \sin(\theta_0)- k x_1(b)\cos(\theta_0) - \eta k x_1'(b)  \cos(\theta_0) \right) .
\end{split}
\end{equation}
To first approximation, $\xi = \eps x_1 \sin(\theta_0)$ and $\eta = \eps z_1 \cos(\theta_0)$. Then
\begin{equation}
    \frac{d\xi}{dt}\approx (c\ssup{0}-U\ssup{0})\Big(-\eps k x_1 \cos(\theta_0)  + \eps^2(x_1^2 k^2 \sin^2(\theta_0)-  k x_1'z_1\cos^2(\theta_0) )\Big).
\end{equation}
Using the double angle identity, 
\begin{equation}\label{eq:xi-approx-2}
\frac{d\xi}{dt}\approx  - \eps k x_1 (c\ssup{0}-U\ssup{0})\cos(\theta_0)
+ \frac{1}{2}\eps^2k(c\ssup{0}-U\ssup{0})\left((k x_1^2 - x_1'z_1)- (x_1'z_1 + kx_1^2)\cos(2\theta_0)\right).
\end{equation}
Using mass continuity \eqref{eq:mass-cont-oeps}, \eqref{eq:xi-approx-2} becomes 
\begin{equation}
\frac{d\xi}{dt}\approx  - \eps k x_1 (c\ssup{0}-U\ssup{0})\cos(\theta_0)
+ \frac{1}{2}\eps^2 (c\ssup{0}-U\ssup{0})\Big(((z_1')^2+z_1''z_1)-  ((z_1')^2-z_1''z_1)\cos(2\theta_0)\Big)
\end{equation}
Integrating in time, we obtain that the first-order portion together with the new second-order portion is
\begin{equation}
\xi = \eps x_1 \sin(\theta_0) + \frac{1}{2}\eps^2(c\ssup{0}-U\ssup{0})((z_1')^2+z_1''z_1)t.
\end{equation}
Now repeating the same procedure for $\eta$, we have 
\begin{equation}\label{eq:eta-1}
\begin{split}
    \frac{d \eta}{d t} = v &\approx \eps k z_1(b+\eta) (c\ssup{0}-U\ssup{0}(b+\eta))\sin(\theta_0) \\ 
    &\approx \eps k (z_1(b) + z_1'(b) \eta)(c\ssup{0}-U\ssup{0}(b)) (\sin(\theta_0) + \xi k \cos(\theta_0)) \\ 
    &\approx \eps k z_1(b)(c\ssup{0}-U\ssup{0}(b)) \sin(\theta_0) + \eps k^2  z_1(b) \xi \cos(\theta_0)+ \eps k z_1'(b) \eta (c\ssup{0}-U\ssup{0}) \sin(\theta_0)\\
    &= \eps k z_1(c\ssup{0}-U\ssup{0})\sin(\theta_0) + \eps^2(c\ssup{0}-U\ssup{0})\Big(k^2x_1z_1 + kz_1'z_1 \Big)\sin(\theta_0)\cos(\theta_0) 
\end{split}
\end{equation}
Since $-kx_1=z_1'$ \eqref{eq:mass-cont-oeps}, the order $\eps^2$ term in \eqref{eq:eta-1} reduces to zero. Then, upon integrating in time, there is no additional contribution at second order to $\eta$, and simply 
\begin{equation}
\eta = \eps x_1 \cos(\theta_0).
\end{equation}
Thus, we can identify $U\ssup{2}$ as $\frac{1}{2}(c\ssup{0}-U\ssup{0})((z_1')^2+z_1''z_1)$. 
Note that one could have alternatively completed this derivation in Eulerian coordinates, as in the original derivation, and used the relationships \eqref{eq:inverted-a-linear} and \eqref{eq:inverted-b-linear} to obtain the final result. This would be possible because one would only account for the leading order portions (which includes the mean flow in $x$) and not the higher order oscillatory terms when doing so, since the additional impact of those (\emph{e.g.}, oscillatory) terms would ultimately only yield $O(\eps^3)$ terms.  
\section{Derivation of alternative form of the pressure}\label{appE}
To derive \eqref{eq:pressure-alternate}, we start with the full Lagrangian momentum equations, \eqref{eq:euler-in-lagrangian-x} and \eqref{eq:euler-in-lagrangian-z}. Multiplying \eqref{eq:euler-in-lagrangian-x} by $x_a$ and \eqref{eq:euler-in-lagrangian-z} by $z_a$, and adding, we obtain
\begin{equation}
    \jac \ddot{x} x_a + \frac1{\rho_0}(p_az_bx_a)+\jac \ddot{z}z_a + \frac{1}{\rho_0}(- p_ax_bz_a + \rho_0 \jac g z_a)=0 
\end{equation}
Dividing both sides by $\jac$,
\begin{equation}
\ddot{x}x_a + \ddot{z}z_a + gz_a + \frac{1}{\rho_0}p_a=0
\end{equation}
Now we use the identities $\ddot{x}=-(c-U)\dot{x}_a$ and $\ddot{z}=-(c-U)\dot{z}_a$ (derived from \eqref{def:monochromatic}) to obtain 
\begin{equation}
    - (c-U)\dot{x}_ax_a - (c-U)\dot{z}_az_a + gz_a + \frac{1}{\rho_0}p_a=0.
\end{equation}
Thus 
\begin{equation}
    \frac{1}{\rho_0}p_a = -gz_a + (c-U)(\dot{x}_ax_a + \dot{z}_az_a).
\end{equation}
Integrating in $a$, we obtain 
\begin{equation}\label{eq:appendix-alternate-pressure}
    p = \rho_0(- g z +(c-U)\frac{1}{2}(\dot{x}^2+\dot{z}^2) + f(b,t)),
\end{equation}
where $f(b,t)$ is a constant of integration that can be determined by the value of the pressure at the surface, and setting $\langle p (b=0)\rangle$ where $p$ is evaluated according to \eqref{eq:appendix-alternate-pressure}. 
\par The momentum budget \eqref{eq:momentum-budget} is then derived using integration by parts twice, various identities relating derivatives in $a$ and $t$ derived from \eqref{def:monochromatic}, and \eqref{eq:appendix-alternate-pressure}.
\end{appen}
\clearpage
\bibliographystyle{jfm}
\bibliography{jfm}

\begin{thebibliography}{40}
\expandafter\ifx\csname natexlab\endcsname\relax\def\natexlab#1{#1}\fi
\def\au#1{#1} \def\ed#1{#1} \def\yr#1{#1}\def\at#1{#1}\def\jt#1{\textit{#1}} \def\bt#1{#1}\def\bvol#1{\textbf{#1}} \def\vol#1{#1} \def\pg#1{#1} \def\publ#1{#1}\def\arxiv#1{#1}\def\org#1{#1}\def\st#1{\textit{#1}}

\bibitem[Al-Zanaidi \& Hui(1984)]{al1984turbulent}
{\sc \au{Al-Zanaidi, MA} \& \au{Hui, WH}} \yr{1984}  \at{Turbulent airflow over water waves-a numerical study}.  \jt{Journal of Fluid Mechanics}  \bvol{148},  \pg{225--246}.

\bibitem[Andrews \& McIntyre(1978)]{andrews1978exact}
{\sc \au{Andrews, D.G.} \& \au{McIntyre, M.E.}} \yr{1978}  \at{An exact theory of nonlinear waves on a {L}agrangian-mean flow}.  \jt{Journal of Fluid Mechanics}  \bvol{89}~(4),  \pg{609--646}.

\bibitem[Belcher {\em et~al.\/}(2012)Belcher, Grant, Hanley, Fox-Kemper, Van~Roekel, Sullivan, Large, Brown, Hines, Calvert {\em et~al.\/}]{belcher2012}
{\sc \au{Belcher, Stephen~E}, \au{Grant, Alan~LM}, \au{Hanley, Kirsty~E}, \au{Fox-Kemper, Baylor}, \au{Van~Roekel, Luke}, \au{Sullivan, Peter~P}, \au{Large, William~G}, \au{Brown, Andy}, \au{Hines, Adrian}, \au{Calvert, Daley} \& \au{others}} \yr{2012}  \at{A global perspective on {L}angmuir turbulence in the ocean surface boundary layer}.  \jt{Geophysical Research Letters}  \bvol{39}~(18).

\bibitem[Benjamin(1959)]{benjamin1959shearing}
{\sc \au{Benjamin, T~Brooke}} \yr{1959}  \at{Shearing flow over a wavy boundary}.  \jt{Journal of Fluid Mechanics}  \bvol{6}~(2),  \pg{161--205}.

\bibitem[Bennett(2006)]{bennett2006}
{\sc \au{Bennett, Andrew}} \yr{2006} {\em Lagrangian fluid dynamics\/}.  \publ{Cambridge University Press}.

\bibitem[Blaser {\em et~al.\/}(2024)Blaser, Benamran, Villas~Bôas, Lenain \& Pizzo]{Blaser2024}
{\sc \au{Blaser, Aidan}, \au{Benamran, Raphaël}, \au{Villas~Bôas, Ana~B}, \au{Lenain, Luc} \& \au{Pizzo, Nick}} \yr{2024}  \at{Momentum, energy and vorticity balances in deep-water surface gravity waves}.  \jt{Journal of Fluid Mechanics}  \bvol{997}~(A55),  \pg{A55}.

\bibitem[Buckley {\em et~al.\/}(2020)Buckley, Veron \& Yousefi]{buckley2020surface}
{\sc \au{Buckley, MP}, \au{Veron, F} \& \au{Yousefi, K}} \yr{2020}  \at{Surface viscous stress over wind-driven waves with intermittent airflow separation}.  \jt{Journal of Fluid Mechanics}  \bvol{905},  \pg{A31}.

\bibitem[Buckley \& Veron(2016)]{buckley2016}
{\sc \au{Buckley, Marc~P} \& \au{Veron, Fabrice}} \yr{2016}  \at{Structure of the airflow above surface waves}.  \jt{Journal of Physical Oceanography}  \bvol{46}~(5),  \pg{1377--1397}.

\bibitem[Cao \& Shen(2021)]{cao2021numerical}
{\sc \au{Cao, Tao} \& \au{Shen, Lian}} \yr{2021}  \at{A numerical and theoretical study of wind over fast-propagating water waves}.  \jt{Journal of Fluid Mechanics}  \bvol{919},  \pg{A38}.

\bibitem[Clamond(2007)]{clamond2007lagrangian}
{\sc \au{Clamond, Didier}} \yr{2007}  \at{On the {L}agrangian description of steady surface gravity waves}.  \jt{Journal of Fluid Mechanics}  \bvol{589},  \pg{433--454}.

\bibitem[Drennan {\em et~al.\/}(2003)Drennan, Graber, Hauser \& Quentin]{Drennan2003-ij}
{\sc \au{Drennan, William~M}, \au{Graber, Hans~C}, \au{Hauser, Danièle} \& \au{Quentin, Céline}} \yr{2003}  \at{On the wave age dependence of wind stress over pure wind seas}.  \jt{Journal of Geophysical Research: Oceans}  \bvol{108}~(C3).

\bibitem[Grare {\em et~al.\/}(2013)Grare, Lenain \& Melville]{Grare2013}
{\sc \au{Grare, Laurent}, \au{Lenain, Luc} \& \au{Melville, W~Kendall}} \yr{2013}  \at{Wave-coherent airflow and critical layers over ocean waves}.  \jt{Journal of Physical Oceanography}  \bvol{43}~(10),  \pg{2156--2172}.

\bibitem[Gu{\'e}rin {\em et~al.\/}(2019)Gu{\'e}rin, Desmars, Grilli, Ducrozet, Perignon \& Ferrant]{guerin2019}
{\sc \au{Gu{\'e}rin, Charles-Antoine}, \au{Desmars, Nicolas}, \au{Grilli, St{\'e}phan~T}, \au{Ducrozet, Guillaume}, \au{Perignon, Yves} \& \au{Ferrant, Pierre}} \yr{2019}  \at{An improved {L}agrangian model for the time evolution of nonlinear surface waves}.  \jt{Journal of Fluid Mechanics}  \bvol{876},  \pg{527--552}.

\bibitem[Hamlington {\em et~al.\/}(2014)Hamlington, Van~Roekel, Fox-Kemper, Julien \& Chini]{hamlington2014}
{\sc \au{Hamlington, Peter~E}, \au{Van~Roekel, Luke~P}, \au{Fox-Kemper, Baylor}, \au{Julien, Keith} \& \au{Chini, Gregory~P}} \yr{2014}  \at{Langmuir--submesoscale interactions: Descriptive analysis of multiscale frontal spindown simulations}.  \jt{Journal of Physical Oceanography}  \bvol{44}~(9),  \pg{2249--2272}.

\bibitem[Jenkins(1986)]{Jenkins1986-ov}
{\sc \au{Jenkins, Alastair~D}} \yr{1986}  \at{A theory for steady and variable wind-and wave-induced currents}.  \jt{Journal of Physical Oceanography}  \bvol{16}~(8),  \pg{1370--1377}.

\bibitem[Lamb(1924)]{lamb1924hydrodynamics}
{\sc \au{Lamb, Horace}} \yr{1924} {\em Hydrodynamics\/}.  \publ{University Press}.

\bibitem[Lenain {\em et~al.\/}(2023)Lenain, Smeltzer, Pizzo, Freilich, Colosi, Ellingsen, Grare, Peyriere \& Statom]{lenain2023airborne}
{\sc \au{Lenain, Luc}, \au{Smeltzer, Benjamin~K}, \au{Pizzo, Nick}, \au{Freilich, Mara}, \au{Colosi, Luke}, \au{Ellingsen, Simen~{\AA}}, \au{Grare, Laurent}, \au{Peyriere, Hugo} \& \au{Statom, Nick}} \yr{2023}  \at{Airborne remote sensing of upper-ocean and surface properties, currents and their gradients from meso to submesoscales}.  \jt{Geophysical Research Letters}  \bvol{50}~(8).

\bibitem[Li {\em et~al.\/}(2017)Li, Fox-Kemper, Breivik \& Webb]{li2017}
{\sc \au{Li, Qing}, \au{Fox-Kemper, Baylor}, \au{Breivik, {\O}yvind} \& \au{Webb, Adrean}} \yr{2017}  \at{Statistical models of global {L}angmuir mixing}.  \jt{Ocean Modelling}  \bvol{113},  \pg{95--114}.

\bibitem[Lighthill(1962)]{Lighthill1962}
{\sc \au{Lighthill, M.~J.}} \yr{1962}  \at{Physical interpretation of the mathematical theory of wave generation by wind}.  \jt{Journal of Fluid Mechanics}  \bvol{14},  \pg{385--398}.

\bibitem[Longuet-Higgins(1969)]{longuethiggins1969}
{\sc \au{Longuet-Higgins, Michael~Selwyn}} \yr{1969}  \at{A nonlinear mechanism for the generation of sea waves}.  \jt{Proceedings of the Royal Society of London. A. Mathematical and Physical Sciences}  \bvol{311}~(1506),  \pg{371--389}.

\bibitem[Miles(1957)]{miles1957}
{\sc \au{Miles, John~W}} \yr{1957}  \at{On the generation of surface waves by shear flows}.  \jt{Journal of Fluid Mechanics}  \bvol{3}~(2),  \pg{185--204}.

\bibitem[Miles(1959)]{miles1959generation}
{\sc \au{Miles, John~W}} \yr{1959}  \at{On the generation of surface waves by shear flows. {P}art 2}.  \jt{Journal of Fluid Mechanics}  \bvol{6}~(4),  \pg{568--582}.

\bibitem[Miles(1993)]{miles1993surface}
{\sc \au{Miles, John~W}} \yr{1993}  \at{Surface-wave generation revisited}.  \jt{Journal of Fluid Mechanics}  \bvol{256},  \pg{427--441}.

\bibitem[Nayfeh \& Saric(1972)]{nayfeh1972}
{\sc \au{Nayfeh, Ali~Hasan} \& \au{Saric, William~S}} \yr{1972}  \at{Nonlinear waves in a {K}elvin-{H}elmholtz flow}.  \jt{Journal of Fluid Mechanics}  \bvol{55}~(2),  \pg{311--327}.

\bibitem[Nouguier {\em et~al.\/}(2015)Nouguier, Chapron \& Gu{\'e}rin]{nouguier2015}
{\sc \au{Nouguier, Fr{\'e}d{\'e}ric}, \au{Chapron, Bertrand} \& \au{Gu{\'e}rin, Charles-Antoine}} \yr{2015}  \at{Second-order {L}agrangian description of tri-dimensional gravity wave interactions}.  \jt{Journal of Fluid Mechanics}  \bvol{772},  \pg{165--196}.

\bibitem[Peirson \& Garcia(2008)]{peirson-garcia2008}
{\sc \au{Peirson, William~L} \& \au{Garcia, Andrew~W}} \yr{2008}  \at{On the wind-induced growth of slow water waves of finite steepness}.  \jt{Journal of Fluid Mechanics}  \bvol{608},  \pg{243--274}.

\bibitem[Pizzo {\em et~al.\/}(2023)Pizzo, Lenain, R{\o}mcke, Ellingsen \& Smeltzer]{pizzo2023}
{\sc \au{Pizzo, Nick}, \au{Lenain, Luc}, \au{R{\o}mcke, Olav}, \au{Ellingsen, Simen~{\AA}} \& \au{Smeltzer, Benjamin~K}} \yr{2023}  \at{The role of {L}agrangian drift in the geometry, kinematics and dynamics of surface waves}.  \jt{Journal of Fluid Mechanics}  \bvol{954},  \pg{R4}.

\bibitem[Plant(1982)]{Plant1982-ty}
{\sc \au{Plant, William~J}} \yr{1982}  \at{A relationship between wind stress and wave slope}.  \jt{Journal of Geophysical Research}  \bvol{87}~(C3),  \pg{1961--1967}.

\bibitem[Rai {\em et~al.\/}(2021)Rai, Hecht, Maltrud \& Aluie]{rai2021}
{\sc \au{Rai, Shikhar}, \au{Hecht, Matthew}, \au{Maltrud, Matthew} \& \au{Aluie, Hussein}} \yr{2021}  \at{Scale of oceanic eddy killing by wind from global satellite observations}.  \jt{Science Advances}  \bvol{7}~(28).

\bibitem[Riley {\em et~al.\/}(1982)Riley, Donelan \& Hui]{Riley1982}
{\sc \au{Riley, D~S}, \au{Donelan, M~A} \& \au{Hui, W~H}} \yr{1982}  \at{An extended {M}iles' theory for wave generation by wind}.  \jt{Boundary Layer Meteorology}  \bvol{22}~(2),  \pg{209--225}.

\bibitem[Stokes(1847)]{stokes1847}
{\sc \au{Stokes, George~Gabriel}} \yr{1847}  \at{On the theory of oscillatory waves}.  \jt{Transactions of the Cambridge Philosophical Society}  \bvol{8},  \pg{441--455}.

\bibitem[Sullivan \& McWilliams(2010)]{sullivan-mcwilliams2010}
{\sc \au{Sullivan, Peter~P} \& \au{McWilliams, James~C}} \yr{2010}  \at{Dynamics of winds and currents coupled to surface waves}.  \jt{Annual Review of Fluid Mechanics}  \bvol{42}~(1),  \pg{19--42}.

\bibitem[Sullivan {\em et~al.\/}(2014)Sullivan, McWilliams \& Patton]{sullivan2014}
{\sc \au{Sullivan, Peter~P}, \au{McWilliams, James~C} \& \au{Patton, Edward~G}} \yr{2014}  \at{Large-eddy simulation of marine atmospheric boundary layers above a spectrum of moving waves}.  \jt{Journal of the Atmospheric Sciences}  \bvol{71}~(11),  \pg{4001--4027}.

\bibitem[Thorpe(2004)]{thorpe2004}
{\sc \au{Thorpe, SA}} \yr{2004}  \at{Langmuir circulation}.  \jt{Annual Review of Fluid Mechanics}  \bvol{36}~(1),  \pg{55--79}.

\bibitem[Valenzuela(1976)]{valenzuela1976growth}
{\sc \au{Valenzuela, GR}} \yr{1976}  \at{The growth of gravity-capillary waves in a coupled shear flow}.  \jt{Journal of Fluid Mechanics}  \bvol{76}~(2),  \pg{229--250}.

\bibitem[Van~Duin \& Janssen(1992)]{van1992analytic}
{\sc \au{Van~Duin, Cornelis~A} \& \au{Janssen, Peter~AEM}} \yr{1992}  \at{An analytic model of the generation of surface gravity waves by turbulent air flow}.  \jt{Journal of Fluid Mechanics}  \bvol{236},  \pg{197--215}.

\bibitem[Weber(1983)]{weber1983}
{\sc \au{Weber, Jan~Erik}} \yr{1983}  \at{Steady wind-and wave-induced currents in the open ocean}.  \jt{Journal of Physical Oceanography}  \bvol{13}~(3),  \pg{524--530}.

\bibitem[Weber(2003)]{weber2003wave}
{\sc \au{Weber, Jan~Erik}} \yr{2003}  \at{Wave-induced mass transport in the oceanic surface layer}.  \jt{Journal of Physical Oceanography}  \bvol{33}~(12),  \pg{2527--2533}.

\bibitem[Weber \& Melsom(1993)]{Weber1993-fy}
{\sc \au{Weber, Jan~Erik} \& \au{Melsom, Arne}} \yr{1993}  \at{Transient ocean currents induced by wind and growing waves}.  \jt{Journal of Physical Oceanography}  \bvol{23}~(2),  \pg{193--206}.

\bibitem[Young \& Wolfe(2014)]{young2014}
{\sc \au{Young, WR} \& \au{Wolfe, CL}} \yr{2014}  \at{Generation of surface waves by shear-flow instability}.  \jt{Journal of Fluid Mechanics}  \bvol{739},  \pg{276--307}.

\end{thebibliography}

\end{document}